\documentclass[sigconf]{acmart}

\renewcommand\footnotetextcopyrightpermission[1]{} % removes footnote with conference information in first column

\AtBeginDocument{%
  }

\setcopyright{none}
\acmConference[]{}{}{}
\usepackage[utf8]{inputenc}
\usepackage{amsthm}
\usepackage{amsmath}
\usepackage{xcolor}
\usepackage{graphicx}
\usepackage{wrapfig}
\usepackage{hyperref}
\usepackage{xspace}
\usepackage{cleveref}
\usepackage[bibliography=common]{apxproof}

\newtheoremrep{theorem}{Theorem}
\newtheoremrep{lemma}[theorem]{Lemma}
\newtheoremrep{definition}[theorem]{Definition}
\newtheoremrep{proposition}[theorem]{Proposition}
\newtheoremrep{corollary}[theorem]{Corollary}
\newtheoremrep{claim}[theorem]{Claim}
\newtheoremrep{note}[theorem]{Note}

\newcommand\var[1]{{\scriptsize #1}}

\usepackage{tikz}
\usetikzlibrary{arrows}
\tikzset{>=stealth}
\tikzset{every path/.style={line width=.3mm,shorten >=1pt,shorten <=1pt}}
\tikzset{every node/.style={minimum size=1pt,fill,circle,inner sep=1pt}}
\tikzstyle{nodevar} = [fill=none, inner sep=-1pt,circle=none]
\tikzstyle{mynodeA} = []
\tikzstyle{mynodeAA} = []
\tikzstyle{mynode} = []
\graphicspath{ {./images/} }
\newcommand{\qdiamond}{\node[nodevar] (A) at (0,0) {{\scriptsize 1}};
\node[nodevar]  (B) at (.5,-.5) {{\scriptsize 4}};
\node[nodevar] (C) at (0,-1) {{\scriptsize 3}};
\node[nodevar] (D) at (-.5,-.5) {{\scriptsize 2}};
\draw[->] (A) -- (D);\draw[->] (D) -- (C);\draw[->] (A) -- (B);}
\newcommand{\core}{
\node[nodevar] (A) at (0,0) {{\scriptsize 1}};
\node[nodevar] (B) at (.7,0) {{\scriptsize 2}};
\node[nodevar] (C) at (1.4,0) {{\scriptsize 3}};
\draw[->] (A) -- (B);
\draw[->] (B) -- (C);
\draw[color=red] (B) circle (4pt);}

\newcommand\blowfish{
\node[nodevar] (A) at (0,0) {\var{1}};
\node[nodevar] (B) at (.5,-.5) {\var{8}};
\node[nodevar] (C) at (1,-1) {\var{7}};
\node[nodevar] (D) at (.5,-1.5) {\var{6}};
\node[nodevar] (E) at (0,-2) {\var{5}};
\node[nodevar] (BB) at (-.5,-.5) {\var{2}};
\node[nodevar] (CC) at (-1,-1) {\var{3}};
\node[nodevar] (DD) at (-.5,-1.5) {\var{4}};

\draw[->] (A) -- (B);
\draw[->] (B) -- (C);
\draw[->] (D) -- (C);
\draw[->] (E) -- (D);
\draw[->] (A) -- (BB);
\draw[->] (BB) -- (CC);
\draw[->] (DD) -- (CC);
\draw[->] (E) -- (DD);
\draw[color=red] (BB) circle (5pt);
}
\newcommand\blowfisheasy{
\blowfish
\node[mynode] (S1) at (0,.5) {};
\node[mynode] (S2) at (-1.5,-1) {};
\node[mynode] (S3) at (1,0) {};
\node[mynodeAA] (S4) at (1.6,-1) {};
\node[mynode] (S5) at (-1,-2) {};
\node[mynodeA] (S6) at (.5,-2.5) {};
\node[mynode] (S7) at (-.5,-2.5) {};
}

\newcommand{\size}[1]{\ensuremath{|\!|#1|\!|}}
\newcommand\pair[1]{\langle #1 \rangle}

\newcommand{\hyperclique}{\textsc{Hyperclique}\xspace}
\newcommand{\sparsehyperclique}{\textsc{sHyperclique}\xspace}
\newcommand{\BMM}{\textsc{BMM}\xspace}
\newcommand{\sparseBMM}{\textsc{sBMM}\xspace}
\newcommand{\UTD}{\textsc{VUTD}\xspace}

\begin{document}

%%
%% The "title" command has an optional parameter,
%% allowing the author to define a "short title" to be used in page headers.
\title{Conjunctive Queries With Self-Joins, Towards a Fine-Grained Complexity Analysis}

%%
%% The "author" command and its associated commands are used to define
%% the authors and their affiliations.
%% Of note is the shared affiliation of the first two authors, and the
%% "authornote" and "authornotemark" commands
%% used to denote shared contribution to the research.
\author{Nofar Carmeli}
\affiliation{
\institution{DI ENS, ENS, Université PSL, CNRS, Inria}
\city{Paris}
\country{France}
}

\author{Luc Segoufin}
\affiliation{
\institution{INRIA, ENS Paris, PSL}
\city{Paris}
\country{France}
}

%%
%% The abstract is a short summary of the work to be presented in the
%% article.
\begin{abstract}
Even though query evaluation is a fundamental task in databases, known classifications of conjunctive queries by their fine-grained complexity only apply to queries without self-joins.
We study how self-joins affect enumeration complexity, with the aim of building upon the known results to achieve general classifications.
We do this by examining the extension of two known dichotomies: one with respect to linear delay, and one with respect to constant delay after linear preprocessing.
As this turns out to be an intricate investigation, this paper is structured as an example-driven discussion that initiates this analysis.
We show enumeration algorithms that rely on self-joins to efficiently evaluate queries that otherwise cannot be answered with the same guarantees. Due to these additional tractable cases, the hardness proofs are more complex than the self-join-free case. We show how to harness a known tagging technique to prove hardness of queries with self-joins.
Our study offers sufficient conditions and necessary conditions for tractability and settles the cases of queries of low arity and queries with cyclic cores. Nevertheless, many cases remain open.
\end{abstract}

%%
%% The code below is generated by the tool at http://dl.acm.org/ccs.cfm.
%% Please copy and paste the code instead of the example below.
%%
%\ccsdesc[500]{Theory of computation~Database query processing and optimization (theory)}

%%
%% Keywords. The author(s) should pick words that accurately describe
%% the work being presented. Separate the keywords with commas.
%\keywords{conjunctive query, self-joins, enumeration, fine-grained, complexity, constant delay}

%%
%% This command processes the author and affiliation and title
%% information and builds the first part of the formatted document.
\maketitle

\section{Introduction}

Query evaluation is one of the most central problems in database systems. This task asks to compute the set of solutions to a given query over a given database.
When treating both the database and the query as input, query evaluation is NP-complete~\cite{Chandra1977OptimalIO}. Here, we adopt the data complexity point of view: the complexity is analyzed with respect to the database size, while the query size (which is usually small) is considered constant. In this setting, query answering takes polynomial time. However, not all queries are equally hard as for some queries this polynomial is much smaller than for others. Ideally, we would like to understand, given any specific query, how fast this query can be evaluated. As the number of solutions may be much larger than the input, considering the delay between solutions and not just the total computation time can be meaningful: the user does not have to wait for the entire computation to be done before seeing some of the solutions.
The ideal time guarantee we can hope to achieve is linear time before the first solution (required to read the input and determine whether a first solution exists) and constant time between successive solutions (required to print the output). 

In this paper we focus on joins and sometimes projections, the basic building blocks of queries, expressed as conjunctive queries.
We know that any \emph{acyclic} query with a \emph{free-connex} form  can be answered with linear preprocessing and constant delay. For {acyclic} queries that are not free-connex, we do not have an algorithm with such guarantees,  but we do have algorithms with linear delay (and linear preprocessing).
At this point, it is natural to ask: 
Are free-connex and acyclic queries the only ones for which such efficient algorithms exist or should we keep looking for efficient algorithms for additional queries?

To answer this question, we have conditional lower bounds showing that other queries cannot be answered within these time bounds unless there would be a significant breakthrough in well-studied problems. However, these hardness results were only proved for \emph{self-join-free} conjunctive queries, queries in which each relation appears at most once.
More specifically, if a self-join-free query is acyclic but not free-connex, then the set of all its solutions cannot be computed in time linear in the input and output sizes, assuming Boolean matrices cannot be multiplied in quadratic time~\cite{bdg:dichotomy}. Moreover, one cannot even produce a single solution to a self-join-free cyclic query in linear time, assuming hypercliques cannot be detected in linear time~\cite{bb:thesis}.
The hardness results described in this introduction also assume the hardness of the two above mentioned algorithmic problems.

For years, database researchers were satisfied with these hardness results despite the limitation to self-join-free queries, as self-joins are known to be problematic.
We remark that in some contexts, e.g.~graph databases, self-joins are very common.
It was believed that the above characterization would extend to queries with self-joins and that this restriction is merely a limitation of our lower-bound tools. This extension was even claimed along with a proof sketch~\cite{bb:thesis}. However, Berkholz, Gerhardt, and Schweikardt~\cite{berkholz2020tutorial} recently refuted this claim by showing a cyclic query with self-joins that can be evaluated efficiently.
That is, we now know that queries with self-joins can be easier than self-join-free queries of the same form.
Now that we have an example that shows it is possible, we would like to understand how, and in which cases, self-joins can be used to streamline query evaluation.
Here, we initiate the complexity analysis for the evaluation of conjunctive queries with self-joins. 

To illustrate the intricacies that arise with self-joins, consider the task of finding all occurrences of the following patterns in a node-colored directed graph (where the red circle denotes a red node). Note that these tasks can be phrased as conjunctive queries with self-joins, containing one binary relation describing the edges and one unary relation describing the red nodes.

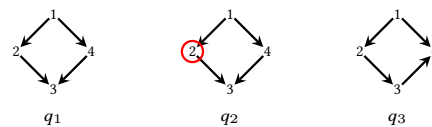
\begin{figure}[ht]
\centering
\begin{tikzpicture}
\qdiamond
\draw[->] (B) -- (C);
\node[nodevar] (Q) at (0,-1.4) {\var{$q_1$}};
\end{tikzpicture}
\hspace{1cm}
\begin{tikzpicture}
\qdiamond
\draw[->] (B) -- (C);
\draw[color=red] (D) circle (4pt);
\node[nodevar] (Q) at (0,-1.4) {\var{$q_2$}};
\end{tikzpicture}
\hspace{1cm}
\begin{tikzpicture}
\qdiamond
\draw[->] (C) -- (B);
\node[nodevar] (Q) at (0,-1.4) {\var{$q_3$}};
\end{tikzpicture}
\caption{Variants of a $4$-cycle with different complexities.}
\label{figure-diamonds}
\end{figure}

We will see that $q_1$, even though cyclic, can be enumerated with constant delay after a linear preprocessing time. Unlike the self-join-free case, adding a unary atom to a query may change its complexity: we will see that $q_2$ can be answered with linear delay but not with constant delay (after only linear preprocessing).  However, for $q_3$ (that differs from $q_1$ by the direction of one edge), we cannot even produce one solution in linear time. This again differs from the self-join-free case, in which the variable order in atoms cannot affect the complexity.

In this paper, we explore the effects of self-joins on query evaluation complexity from an enumeration point of view. The aim is to use known results for the self-join-free case to reason about general queries that may contain self-joins. We show enumeration algorithms that rely on self-joins to efficiently evaluate queries that otherwise cannot be answered with the same guarantees. Due to these additional tractable cases, the hardness proofs also require additional complications compared to the self-join-free case. We show how to harness a known tagging technique to prove hardness of queries with self-joins.

We discuss the difficulty in proving lower bounds for queries with self-joins in \Cref{section-general-hardness} and then define the tagging technique and explain the importance of endomorphism images for its analysis.
\Cref{section-lowarity} uses this technique to establish that self-joins do not change whether a query of arity two or less can be enumerated with constant delay.
In \Cref{sec:cyclic-core}, we show that queries with a cyclic ``core'' are hard in the sense that they cannot be answered with only linear preprocessing, even if they contain self-joins. As a result, we focus on queries with an acyclic core in the rest of the paper. 

Next, we focus on full conjunctive queries, with no projection.
In \Cref{section-linear}, we phrase a sufficient condition and a necessary condition for enumeration with linear delay, but we also give examples of cases where these conditions are not enough.
In \Cref{section-constant}, we identify a class of cyclic queries that can be answered with constant delay (we call these mirror queries). We then provide an example-driven discussion of how simple acyclic additions (unary atoms and spikes) to tractable queries can affect the complexity. We finish with a simple (single-cycle) query of unknown complexity.
Missing details appear in the appendix.

\section{Preliminaries}\label{section-prelim}
A \emph{database} $D$ is a finite relational structure consisting of a relational schema $\sigma$, a finite domain $V$, and for each relational symbol $R$ of $\sigma$ of arity $r$, a subset $R^D$ of $V^r$. A term $R(\bar a)$ where $\bar a \in R^D$ is called a \emph{fact} of $D$. 

In this paper, \emph{query} always refers to a \emph{conjunctive query}, i.e. a conjunction of atoms $R(\bar x)$, where $R \in \sigma$ and $\bar x$ is a tuple of variables of the appropriate arity, with some of the variables existentially quantified.
The variables that are not quantified in $q$ are called the \emph{free variables} of $q$. The \emph{arity} of a query is the number of its free variables. A \emph{Boolean/unary/binary} query is a query of arity 0/1/2 respectively. A  query is \emph{full} if it contains no quantified variables, i.e. all the variables are free. 
A  query has \emph{self-joins} if it contains two atoms with the same relational symbol; otherwise, it is called \emph{self-join-free}. For example, the query $R(x_1,x_2) \land R(x_2,x_3)$ has self-joins while the query $R(x_1,x_2) \land S(x_2,x_3)$ is self-join-free.

We will often specify full conjunctive queries using a colored hypergraph rather than a formula.
Each node represents a variable of the query {and is depicted by the number of the variable or a black dot when this number is not relevant}. Each colored edge represents a binary atom, and each colored circle around a node represents a unary atom. The exact relation names are not depicted, but they are irrelevant to the complexity analysis. As an example, $q_2$ of \Cref{figure-diamonds} defines the query $E(x_1,x_2)\land E(x_2,x_3) \land E(x_1,x_4) \land E(x_4,x_3) \land \textit{Red}(x_2)$.
We sometimes use a big circle to represent a relation of arity three or more (as in $q_6$ of \Cref{figure-untangling-one}). When we do so, the order of the variables in this relation is not important for the example.

A query is \emph{acyclic} if it has a join tree. A \emph{join tree} for the query $q$ is a tree $T$ such that the nodes of $T$ are the atoms of $q$ and, for each variable $x$ of $q$, the nodes of $T$ whose atoms contain $x$ are connected in $T$. An acyclic query $q$ is said to be \emph{free-connex} if we obtain an acyclic query by adding to $q$ a new atom containing all the free variables of $q$. As an example, the acyclic query $\exists x_2 R(x_1,x_2)\land S(x_2,x_3)$ is not free-connex because the query $\exists x_2 R(x_1,x_2)\land S(x_2,x_3)\land T(x_1,x_3)$ is cyclic; however, the full acyclic query $R(x_1,x_2)\land S(x_2,x_3)$ is free-connex because the query $R(x_1,x_2)\land S(x_2,x_3)\land T(x_1,x_2,x_3)$ is acyclic. In fact, it follows from the definition of free-connexity that full acyclic queries are always free-connex. Similarly, Boolean or unary acyclic queries are free-connex as well.

The \emph{solutions} of a query $q$ over a database $D$ are denoted by $q(D)$. They correspond to the set of tuples $\bar a$ such that: there exists an assignment of the elements of the domain of $D$ to the variables of $q$ making every atom of $q$ a fact of $D$, and $\bar a$ is the assignment to the free variables of $q$. When we specify a solution to a query, we assume the free variables are ordered according to their numbering.
The \emph{output size} is the number of tuples in the set $q(D)$. 
Two queries $q$ and $q'$ are said to be \emph{equivalent} if for every database $D$, $q(D)=q'(D)$.

An \emph{endomorphism} of $q$ is a mapping from the variables of $q$ to the variables of $q$ that preserves the atoms of $q$; that is, if $R(\bar x)$ is an atom of $q$ and $\nu$ is an endomorphism of $q$, then $R(\nu(\bar x))$ is also an atom of $q$. An \emph{automorphism} is an injective endomorphism. A query $q$ is said to be \emph{minimal}\footnote{It is called minimal as there is no smaller (with strictly fewer atoms) equivalent query.} if any endomorphism of $q$ that is the identity on the free variables of $q$ is an automorphism of $q$. Notice that every self-join-free query is minimal as the only endomorphism in such a query is the identity mapping. Any query $q$ has an equivalent minimal query, called the \emph{minimal form} of $q$ in what follows.

The problems we consider are parameterized by a query $q$ and take as input a database $D$.
\emph{The evaluation problem} computes the set of solutions $q(D)$.
\emph{The testing problem} asks whether $q(D)$ is empty.
\emph{The enumeration problem} is decomposed into two phases: the preprocessing phase computes a data structure that is used during the enumeration phase for outputting the solutions of $q(D)$ one by one and with no repetition. The maximal time between two consecutive answers during enumeration is called the \emph{delay}.

We use the RAM computation model with uniform addition and multiplication, and we adopt the data complexity point of view.
That is, the constants hidden in the big $O$ notation may depend on the query but not the database. 
The input size, denoted $\size{D}$,
corresponds to the size of a reasonable encoding of the database $D$, linear in the number of database tuples.
Note that an enumeration algorithm with linear preprocessing implies that the testing problem can be solved in linear time. 
A constant delay after linear preprocessing enumeration algorithm implies that the evaluation problem can be solved in linear input plus output time.

Many known enumeration algorithms for database queries conform to a strict definition of linear preprocessing and constant delay where the memory used during preprocessing is linear and the extra memory used after the preprocessing remains bounded by a constant~\cite{DBLP:conf/pods/SchweikardtSV18,DBLP:journals/lmcs/KazanaS19,DBLP:journals/tocl/KazanaS13,bdg:dichotomy,DBLP:journals/corr/abs-1105-3583}.
Our enumeration algorithms use the allowed time in the most general way, without memory restrictions. 
In particular, we assume we can test in constant time whether a given tuple is a fact of the database. It is not clear how to do this test without using perfect hashing, which requires polynomial memory in the size of the input database.
The enumeration phases of our algorithms may also require a total memory of a similar size.
This is required to store each produced solution (or parts of the solution), used later for producing more solutions. As this distinction is important, we will make it explicit when we use this extra memory. This is in particular the case in the proof of the Cheater's Lemma~\cite{UCQs} that we use to eliminate duplicate answers.

\begin{lemma}[(Cheater's Lemma) \cite{UCQs}]\label{cheater-lemma}
Let $q$ be a conjunctive query. If there is an algorithm evaluating $q$ with linear preprocessing time and constant/linear delay, where the number of times each output is produced is bound by a constant,
then there is an enumeration algorithm for $q$ that runs with linear preprocessing time and constant/linear delay, producing the solutions with no repetition.
This new algorithm may use during the enumeration phase a memory of size $O(\size{D}^{|q|})$ over an input database $D$.
\end{lemma}

As usual in this area, our lower bounds assume algorithmic conjectures, described next. 
A hypergraph is $(k-1)$-uniform if every hyperedge is of size $k-1$. A $k$-hyperclique is a set of $k$ nodes such that any subset of these nodes of size $k-1$ forms a hyperedge.
\begin{itemize}
\item \sparseBMM conjectures that two Boolean matrices $A$ and $B$, represented as lists of their non-zero entries, cannot be multiplied in time $O(m)$, where $m$ is the number of non-zero entries in $A$, $B$, and $AB$.
\item \sparsehyperclique conjectures that for all $k\geq 3$, it is not possible to determine the existence of a $k$-hyperclique in a $(k-1)$-uniform hypergraph with $m$ hyperedges in time $O(m)$.
\item \hyperclique{} conjectures that for all $k\geq 3$, it is not possible to determine the existence of a $k$-hyperclique in a $(k-1)$-uniform hypergraph with $n$ nodes in time $O(n^{k-1})$.
\end{itemize}
Notice that \sparsehyperclique implies that we cannot detect a triangle in an undirected graph in linear time. \hyperclique implies the sparse hypotheses \sparsehyperclique and \sparseBMM.

We can now state the known results for query evaluation.
\begin{theorem}[\cite{bdg:dichotomy,bb:thesis}]\label{dichotomy-sjf}
Let $q$ be a self-join-free conjunctive query.
\begin{itemize}
\item If $q$ is cyclic, the testing problem cannot be solved in linear time, assuming \sparsehyperclique.
\item If $q$ is acyclic, the enumeration problem can be solved with linear delay and linear preprocessing.
\item If $q$ is acyclic but not free-connex, the evaluation problem cannot be solved in linear time in the input plus output sizes, assuming \sparseBMM.
\item If $q$ is acyclic and free-connex, the enumeration problem can be solved with constant delay and linear preprocessing. 
\end{itemize}
\end{theorem}
In the above theorem, the upper bounds also hold for queries with self-joins. 
The enumeration algorithms obtained in~\cite{bdg:dichotomy,bb:thesis} use the strict definition of constant delay, where the total memory used during the enumeration phase only adds a constant to the memory used during the preprocessing phase.

\section{Proving Hardness with Self-Joins}\label{section-general-hardness}\label{section-trick}

Since we have a good understanding of the complexity of self-join-free queries, we are interested in comparing the complexity of general conjunctive queries to self-join-free queries with the same shape.
More formally, we associate a self-join-free query $q'$ to any query $q$ by giving distinct names to all relation symbols. As an example, the self-join-free query associated to $R(x_1,x_2) \land R(x_2,x_3)$ is $R_1(x_1,x_2) \land R_2(x_2,x_3)$. Notice that this modification does not affect properties such as acyclicity or free-connexity.

Going from a query $q$ to its associated self-join-free query $q'$ can only make the evaluation task harder. This folklore proposition holds since $q'$ can be used to compute $q$, and it implies that the upper bounds of Theorem~\ref{dichotomy-sjf} also hold for queries with self-joins.

\begin{propositionrep}\label{prop-reduce-to-sjf}
For any conjunctive query $q$, its associated self-join-free query $q'$ is at least as hard as $q$ for {the evaluation problem, the testing problem, and the enumeration problem.}
\end{propositionrep}
\begin{proof}
Given a database $D$ for $q$ we compute a database $D'$ for $q'$ such that the answers of $q$ over $D$ can be reconstructed from the answers of $q'$ over $D'$. The database $D'$ is constructed from $D$ by duplicating each relation $R$ as many times as its number of occurrences in $q$, one for each new symbol used in $q'$ as a replacement of $R$. The reduction can be done in linear time, and $q(D)=q'(D')$.
\end{proof}

When trying to show the other direction, reducing the self-join-free query to its initial query with self-joins, we need more care.
Let us now explain why the lower bounds of \Cref{dichotomy-sjf} only apply to self-join-free queries.

\begin{example}\label{example-sjf-proof-fails}
Consider the self-join-free variant $q_1'$ of $q_1$ from \Cref{figure-diamonds}:
$R_1(x_1,x_2)\land R_2(x_2,x_3) \land R_3(x_1,x_4) \land R_4(x_4,x_3)$.
\Cref{dichotomy-sjf} proves that the ability to determine whether $q_1'$ has answers in linear time would imply finding triangles in an undirected graph in linear time, which is conjectured to be impossible. 
The reduction works as follows: for every edge $(a,b)$ of the input graph with $a<b$,\footnote{To go from an undirected graph to a binary relation we can either order the vertices of the graph as we do here or {construct symmetric relations}.} we add to the database the facts: $R_1(a,b)$, $R_2(a,b)$, $R_3(a,b)$, and $R_4(b,b)$. With this construction, every answer to $q_1'$ is of the form $(a,b,c,c)$ such that $(a,b,c)$ is a triangle.
If we try to use the same technique to prove the hardness of $q_1$, we encounter the problem that we cannot assign different atoms with different relations. In particular, if we add to the database the facts $R(a,b)$ and $R(b,b)$ for every edge $(a,b)$ of the graph, the query $q_1$ will also have solutions corresponding to $2$-paths (where every atom maps to the edge relation and $x_2$ and $x_4$ map to the same value). Thus, we do not get that the query has a solution iff the graph has a triangle.
\end{example}

To prove the hard cases, we need a technique that allows us to assign different relations to different atoms despite the self-joins.
For this reason, we will use variations of the tagging technique described below that aims to assign different variables with different domains by concatenating the variable names to the domain values.

{\bf Tagging technique.}\footnote{This technique was suggested by Bagan, Durand and Grandjean~\cite{bdg:dichotomy} to show the hardness of queries with disequalities. It was later used to show the hardness of unions of conjunctive queries~\cite{UCQs}.
Brault-Baron~\cite[Chapter 4, Lemma 9]{bb:thesis} claimed that it can be used to show that queries with self-joins are as hard as queries without them, but this claim, which was later refuted, was not accompanied by a proof.}
Let $q$ be a minimal conjunctive query and $q'$ be its associated self-join-free query. Let $D'$ be a database for $q'$. We construct a database $D$ for $q$ as follows. Let $R_1(\bar x)$ be an atom of $q'$ associated to the atom $R(\bar x)$ of $q$, where $R$ is of arity $l$. For every fact $R_1(\bar a)$ of $D'$, we add to $D$ the fact $R(\pair{a_1,x_1},\ldots,\pair{a_l,x_l})$.

With this construction, every element of the domain of $D$ is of the form $\pair{a,x}$. We refer to $a$ as the data part and to $x$ as the tag.
Let us now inspect $q(D)$.
Every solution $(\pair{a_1,y_1},\ldots,\pair{a_k,y_k})$ in $q(D)$ is witnessed by an assignment $\mu$ associating an element of $D$ to every variable of $q$. As above, $\mu$ can be split into two parts: a data map associating $a_i$ to $x_i$ whenever $\mu(x_i)=\pair{a_i,y_i}$ and a tagging map associating $y_i$ to $x_i$ in the same cases.
The data part of the solution of $\mu$ is $(a_1,\cdots,a_k)$.
Notice that every solution to $q'(D')$ appears as a data part of a solution of $q(D)$ with an identity tagging map.
For the converse direction, consider a solution of $q(D)$. There are two cases.
\begin{enumerate}
\item If the tagging map $\nu$ is an automorphism, then the data map contains a solution of $q'(D')$ as follows. Consider the mapping $\mu'$ defined by $\mu'(x)=\mu(\nu^{-1}(x))$. It clearly witnesses a solution of $q(D)$ with an identity tagging map whose data part then belongs to  $q'(D')$.
\item If the tagging map is not an automorphism, the data part of the solution may then not correspond to a solution in $q'(D')$.
\end{enumerate}
If $q$ is minimal and the tagging map $\nu$ is the identity on the free variables, then $\nu$ must be an automorphism. Thus, when using the tagging technique, solutions with the identity tagging map are the ones we wish to return in order to answer $q'$, and we need to ensure that there are not too many interfering solutions of the other cases.
We will see that if the query is of low arity, then the number of extra solutions remains small and so they be ignored without incurring a high cost.
For full queries, other solutions of the first case are not a problem: As we are given the assignment to all variables as part of a solution, we can compute $\mu'$ and an answer to $q'$ given any such solution to $q$. Every solution to $q'$ will be obtained a constant number of times this way, once for each automorphism, and these duplicates can be eliminated using the Cheater's Lemma.
Solutions of the second case are the ones we need to beware of. For this technique to result in a useful reduction, we need to limit the number of these extra solutions, and this cannot always be achieved. 
We summarize the discussion above into the following lemma.
\begin{lemma}\label{trick-lemma}
Let $q'$ be the self-join-free query associated with a minimal conjunctive query $q$, and let $D$ be the database for $q$ constructed using the tagging technique from a database $D'$ for $q'$. Then,
$q'(D')$ is precisely the data parts of those solutions to $q(D)$ in which every free variable is assigned itself as a tag. Moreover, if $q$ is full, then each solution to $q(D)$ tagged by an automorphism can be translated in constant time to a solution of $q'(D')$ such that each solution of $q'(D')$ is obtained a constant number of times.
\end{lemma}

The following is an example of the tagging technique and how it can fail to prove hardness due to extra solutions. Later in the paper, we will see how this difficulty can be circumvented in some cases so that the tagging technique works.

\begin{example}
Consider $q_1'$ and $q_1$ from Example~\ref{example-sjf-proof-fails}.
Notice that $q_1$ has two automorphisms: the identity and the one that switches $x_2$ and $x_4$. It also has two other endomorphisms: the one that maps $x_2$ and $x_4$ to $x_2$, and the one that maps them both to $x_4$.
Consider the database $D'$ consisting of the four facts $R_1(a,b)$, $R_2(b,c)$, $R_3(a,d)$ and $R_4(d,c)$.
Then, $q_1'(D')$ consists of one solution $(a,b,c,d)$. The database $D$ constructed by the tagging technique consists of the facts $R(\pair{a,x_1},\pair{b,x_2})$, $R(\pair{b,x_2},\pair{c,x_3})$, $R(\pair{a,x_1},\pair{d,x_4})$, and \allowbreak $R(\pair{d,x_4},\pair{c,x_3})$. 
The solutions $q_1(D)$ contain the solutions of the first case: $(\pair{a,x_1},\pair{b,x_2},\pair{c,x_3},\pair{d,x_4})$ and 
$(\pair{a,x_1}$,$\pair{d,x_4}$, $\pair{c,x_3}$, $\pair{b,x_2})$. In other words, these answers contain two copies of the initial solution $(a,b,c,d)$, one per automorphism of $q_1$. However, $q_1(D)$ also contains the solutions $(\pair{a,x_1}$,$\pair{b,x_2}$, $\pair{c,x_3}$, $\pair{b,x_2})$ and  $(\pair{a,x_1},\pair{d,x_4},\pair{c,x_3},\pair{d,x_4})$, whose data parts correspond to the paths of length two $(a,b,c)$ and $(a,d,c)$.
In this case, the tagging technique may produce too many solutions on general databases (a solution for each $2$-path), and so this trick cannot be used for showing the hardness of $q_1$. In fact, we will see in \cref{section-constant} that $q_1$ can be answered with constant delay and is therefore easier than its self-join-free form.
\end{example}

\section{Queries of Low Arity}\label{section-lowarity}

We use the tagging from the previous section to show that self-joins do not affect the tractable cases for constant delay enumeration of conjunctive queries with few free variables.

For Boolean and unary queries, the number of solutions is at most linear in the database size. If we could compute the entire solution set in linear time then constant delay enumeration easily follows. This situation is characterized by the following result.

\begin{theoremrep}\label{thm-unary}
Let $q$ be a minimal conjunctive query of arity at most one. Then, its solutions can be computed in linear time iff it is acyclic, assuming \sparsehyperclique.\footnote{In fact, the proof of the unary case can be applied to any minimal conjunctive query in which one atom contains all free variables.}
\end{theoremrep}
\begin{proof}
If $q$ is acyclic, since it is Boolean or unary, it is free-connex, and so it can be answered with constant delay after linear preprocessing time by Theorem~\ref{dichotomy-sjf}. This finds all solutions in linear time.
Assume now that $q$ is cyclic.

To show that $q$ is hard, we use a reduction based on the tagging technique. 
Let $q'$ be the self-join-free query associated to $q$. As $q'$ is cyclic, its set of solutions cannot be computed in linear time by Theorem~\ref{dichotomy-sjf}, assuming \sparsehyperclique. Let $D'$ be a database for $q'$ and let $D$ be the database constructed from $D'$ using the tagging technique.

If $q$ is Boolean, since it is also minimal, every endomorphism is an automorphism, so by Lemma~\ref{trick-lemma}, $q(D)$ is true iff $q'(D')$ is true too.

Otherwise, $q$ is unary, and let $x$ be the free variable of $q$. From Lemma~\ref{trick-lemma}, we get that 
$q'(D')=\{a\mid\pair{a,x}\in q(D)\}$.
As $q$ is unary, the total size of $q(D)$ is linear in the size of $D$ and so also in the size of $D'$. Hence, if we could compute all the solutions $q(D)$ in linear time, we could compute all the solutions to $q'(D')$ at the same time simply by ignoring the solutions in $q(D)$ that are not tagged by $x$.
\end{proof}
\begin{proofsketch}
{
If $q$ is cyclic then its associated self-join-free query $q'$ is also cyclic and therefore hard by \cref{dichotomy-sjf}. We reduce the evaluation of $q'$ to the evaluation of $q$ using the tagging technique and extract the solutions to $q'$ as in \cref{trick-lemma}. As $q$ is at most unary, it has a linear number of solutions, and so we can ignore the extra solutions at a low cost.
}
\end{proofsketch}

We remark that, while \Cref{thm-unary} implies that cyclic unary queries cannot be enumerated with constant delay, it does not imply the same for linear delay. In fact, the cyclic unary query $q_4(x)$ defined as $\exists yuv E(u,y) \land E(y,v) \land E(u,x) \land E(x,v) \land P(y)$ can be enumerated with linear delay.
It remains open to characterize the unary queries with respect to linear delay.

\begin{toappendix}
{\bf Missing details for $q_4$.}
The enumeration of $q_4$ in linear delay can be done as follows. We first precompute the set $S$ of all elements $a$ such that $P(a)$ is a fact of the database and $a$ has incoming and outgoing edges. Notice that this is an acyclic property and therefore $S$ can be computed in linear time. Notice also that any element of $S$ is a solution to $q$. For every element $a$ of $S$ we do the following. First, we consider all edges $E(b,c)$. If $E(b,a)$ is an edge, then we store $c$ in a temporary table $T$. Next, we start a new scan of all edges $E(b,c)$. If $E(a,c)$ is an edge and $b\in T$, then we can safely output $b$. For each $a\in S$, at least one solution is printed, namely $a$, and the delay between checking two elements of $S$ is linear.
\end{toappendix}

We show a similar result to that of \Cref{thm-unary} for binary queries using the non-sparse version of the hypothesis.

\begin{theoremrep}\label{theorem-binary}
Let $q$ be a minimal binary conjunctive query. 
If $q$ is not acyclic free-connex, then there is no linear preprocessing and constant delay enumeration algorithm for answering $q$, assuming \hyperclique.
\end{theoremrep}
\begin{proof}
Let $q$ be a minimal binary query, and let $q'$ be its associated self-join-free query. 
Assume $q$ is acyclic but not free-connex. See the paper body for the proof in the cyclic case. 
We use the following conjecture which is implied by \hyperclique.
\BMM conjectures that two $n\times n$ Boolean matrices $A$ and $B$ cannot be multiplied in time $O(n^2)$.
We show that we cannot compute $q(D)$ in time linear in the input and output sizes, assuming \BMM.
To see this, we need to recall how the same result was shown in the self-join-free case. 
Given two $n \times n$ matrices, there is a construction that encodes their multiplication task into any acyclic non-free-connex query using $n$ domain values~\cite{bdg:dichotomy}. If the query could be solved in time linear in the input and output sizes, this construction would multiply the matrices in $O(n^2)$ time.
We use the same construction to define a database $D'$ for $q'$.
Using the tagging technique to produce $D$ from $D'$, $q(D)$ contains $q'(D')$ and possibly additional answers.
But, as the query is binary, the size of $q(D)$ is at most $O(n^2)$. Hence, if we could compute $q(D)$ in linear time, we would compute $q'(D')$ with an overhead of no more than $O(n^2)$ by ignoring the unwanted solutions, contradicting \BMM.
\end{proof}
\begin{proofsketch}
Let $q$ be a minimal binary query and $q'$ be its associated self-join-free query. 
Assume that $q$ is cyclic. In the self-join-free case, a known construction~\cite{bb:thesis} encodes the hyperclique detection problem into solving any cyclic query over a database with $n$ domain values, where $n$ is the number of vertices in the hypergraph. Hence, a solution to $q'$ cannot be computed in linear time, assuming \hyperclique. Using the tagging technique, the number of extra solutions to $q(D)$ is bounded by $O(n^2)$. So, for any $k\geq 3$, the time necessary to ignore the unwanted solutions falls within the allowed time of $O(n^{k-1})$ for testing the existence of a $k$-hyperclique. The acyclic non-free-connex case is similar.
\end{proofsketch}

\section{Queries With a Cyclic Core}\label{sec:cyclic-core}

{Given a conjunctive query $q$ with free variables $\bar x$, let $p$ be the minimal form of the Boolean query $\exists \bar x q$. We denote by the \emph{core} of $q$ the query constructed from $p$ by removing all quantifications. Notice that the core of $q$ is a full query that may not be equivalent to $q$. Notice also that an acyclic query has an acyclic core but the converse is not necessarily true: the core of the cyclic query $q_1$ of Figure~\ref{figure-diamonds} is acyclic.}  
Observe that a query has a solution iff its core has a solution. Hence the following result is an immediate consequence of \Cref{thm-unary}.

\begin{theoremrep}\label{min-bool-cyclic}
It is possible to test in linear time whether a conjunctive query has a solution iff it has an acyclic core, assuming \sparsehyperclique.
\end{theoremrep}
\begin{proof}
Let $q$ be a conjunctive query with free variables $\bar x$, and let $p$ be the Boolean version of the core of $q$. Notice that $p$ has a solution iff $q$ has a solution.
If $p$ is acyclic, then we can test whether it has a solution in linear time using the Yannakakis algorithm~\cite{DBLP:conf/vldb/Yannakakis81}. 
Otherwise, $p$ is a minimal cyclic Boolean query. By \Cref{thm-unary}, $p$ cannot be tested in linear time assuming \sparsehyperclique. Hence, one cannot test in linear time whether $q$ has a solution.
\end{proof}

\begin{example}\label{example-cyclic-core}
Consider the full conjunctive query $q_3$ depicted by \Cref{figure-diamonds}. The core of the query is the query itself and is therefore cyclic. It follows from \Cref{min-bool-cyclic} that the query is hard, i.e. we cannot test whether it has a solution in linear time.
\end{example}

\section{Full Conjunctive Queries and Linear Delay}\label{section-linear}

Assuming \sparsehyperclique, acyclic self-join-free queries can be enumerated with linear delay (after linear preprocessing) while cyclic self-join-free queries cannot.
We now ask how self-joins affect this classification. Following \cref{prop-reduce-to-sjf} and 
\Cref{min-bool-cyclic}, it
remains to handle cyclic queries with an acyclic core.
In this section, we begin this investigation by considering only full queries of this form.
We start with a simple example showing that some cyclic queries with self-joins can be enumerated with linear delay.

\begin{example}\label{example-acycliccore-linear}
Consider the full query $q_2$ of \Cref{figure-diamonds}. It can be enumerated with linear delay as follows. We enumerate the solutions of the core of $q_2$, depicted by \begin{tikzpicture}
\core
\end{tikzpicture}. As the core is acyclic and free-connex, by \Cref{dichotomy-sjf}, this can be done with constant delay after linear preprocessing. For each solution $(a,b,c)$ of the core, we try all elements $d$ of the database and test, in constant time, whether $d$ is connected to $a$ and $c$ as specified by $q_2$. If it is, we output the solution $(a,b,c,d)$. Remark that any solution $(a,b,c)$ to the core gives rise to a solution $(a,b,c,b)$ to $q_2$. Thus, at least one solution to $q_2$ is printed for each solution to the core, and so the delay is at most linear.
\end{example}

\subsection{Sufficient condition}\label{section-linear-easy}

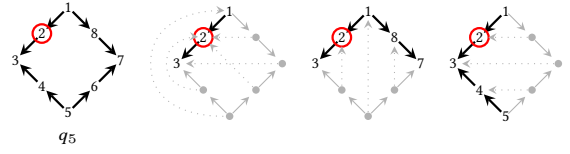
\begin{figure}[ht]
\centering
\begin{tikzpicture}[scale=.7]
\blowfish
\node[nodevar] (Q) at (0,-2.5) {\var{$q_5$}};
\end{tikzpicture}
\hspace{0.2cm}
\begin{tikzpicture}[scale=.7]
\node[nodevar] (A) at (0,0) {\var{1}};
\node[nodevar] (BB) at (-.5,-.5) {\var{2}};
\node[nodevar] (CC) at (-1,-1) {\var{3}};
\draw[->] (A) -- (BB);
\draw[->] (BB) -- (CC);
\draw[color=red] (BB) circle (5pt);

\tikzset{every path/.style={color=black!30,shorten >=2pt}}
\node (E) at (0,-2) {};
\node (B) at (.5,-.5) {};
\node (C) at (1,-1) {};
\node (D) at (.5,-1.5) {};
\node (DD) at (-.5,-1.5) {};
\draw[->] (E) -- (DD);
\draw[->] (A) -- (B);
\draw[->] (B) -- (C);
\draw[->] (D) -- (C);
\draw[->] (E) -- (D);
\draw[->] (DD) -- (CC);
\draw[dotted,->] (B) -- (BB);
\draw[dotted,->] (C) -- (CC);
\draw[dotted,->] (D) -- (BB);
\draw[dotted,->] plot [smooth,tension=2] coordinates {(E) (-1.5,-1) (A)};
\draw[dotted,->] plot [smooth,tension=2] coordinates {(DD) (-1.2,-1) (BB)};
\node[nodevar] (Q) at (0,-2.5) {};
\end{tikzpicture}
\hspace{0.2cm}
\begin{tikzpicture}[scale=.7]
\node[nodevar] (A) at (0,0) {\var{1}};
\node[nodevar] (B) at (.5,-.5) {\var{8}};
\node[nodevar] (C) at (1,-1) {\var{7}};
\node[nodevar] (BB) at (-.5,-.5) {\var{2}};
\node[nodevar] (CC) at (-1,-1) {\var{3}};
\draw[->] (A) -- (B);
\draw[->] (B) -- (C);
\draw[->] (A) -- (BB);
\draw[->] (BB) -- (CC);
\draw[color=red] (BB) circle (5pt);
\tikzset{every path/.style={color=black!30,shorten >=2pt}}
\node (D) at (.5,-1.5) {};
\node (E) at (0,-2) {};
\node (DD) at (-.5,-1.5) {};
\draw[->] (D) -- (C);
\draw[->] (E) -- (D);
\draw[->] (DD) -- (CC);
\draw[->] (E) -- (DD);
\draw[dotted,->] (DD) -- (BB);
\draw[dotted,->] (D) -- (B);
\draw[dotted,->] (E) -- (A);
\node[nodevar] (Q) at (0,-2.5) {};

\end{tikzpicture} \hspace{0.2cm} \begin{tikzpicture}[scale=.7]
\node[nodevar] (A) at (0,0) {\var{1}};
\node[nodevar] (E) at (0,-2) {\var{5}};
\node[nodevar] (BB) at (-.5,-.5) {\var{2}};
\node[nodevar] (CC) at (-1,-1) {\var{3}};
\node[nodevar] (DD) at (-.5,-1.5) {\var{4}};
\node[nodevar] (Q) at (0,-2.5) {};

\draw[->] (A) -- (BB);
\draw[->] (BB) -- (CC);
\draw[->] (DD) -- (CC);
\draw[->] (E) -- (DD);
\draw[color=red] (BB) circle (5pt);

\tikzset{every path/.style={color=black!30,shorten >=2pt}}
\node (B) at (.5,-.5) {};
\node (C) at (1,-1) {};
\node (D) at (.5,-1.5) {};
\draw[->] (A) -- (B);
\draw[->] (B) -- (C);
\draw[->] (D) -- (C);
\draw[->] (E) -- (D);
\draw[dotted,->] (B) -- (BB);
\draw[dotted,->] (C) -- (CC);
\draw[dotted,->] (D) -- (DD);

\end{tikzpicture}
\caption{
The query $q_5$ and its images. We depict images in dark and the corresponding endomorphisms in dotted gray. Self-loops are not depicted for readability.}\label{figure-blowfish}
\end{figure}

We now present a sufficient condition for enumeration with linear delay that builds on two key concepts: \emph{image} and \emph{untangling}.
Consider an endomorphism $\nu$ of a full query $q$. The full query consisting of all atoms $R(\nu(\vec{x}))$ such that $R(\vec{x})$ is an atom of $q$ is called an \emph{image} of $q$.
\Cref{figure-blowfish} depicts examples of images.
By definition, the core of $q$ is an image.
Another trivial image is the query $q$ itself.

The key property of images is that any solution to the image yields a solution to the full query: Simply instantiate any variable of the query by the value of its image. As an example, for the query $q_5$ depicted in \Cref{figure-blowfish}. Every solution $(a,b,c)$ of the core yields the solution $(a,b,c,b,a,b,c,b)$ of $q_5$, and every solution $(a,b,c,d,e)$ of the rightmost depicted image yields the solution $(a,b,c,d,e,d,c,b)$ of $q_5$. Hence, if the image is acyclic, we can use it to generate many ``simple'' solutions in linear time and constant delay, giving us some time to compute more complicated solutions.

\newcommand\projectedq[2]{{#1}|_{#2}}

The \emph{restriction} of a conjunctive query $q$ to a set of its variables $S$, denoted $\projectedq{q}{S}$, is the query obtained from the self-join-free query associated to $q$ by removing all variables not in $S$ from its atoms. Notice that the arities of the relations may change.
An \emph{untangling} of $q$ is a sequence of images $I_0,\ldots,I_k$ of $q$ such that: $I_0=q$, $I_k$ is acyclic, and for all $0\le i<k$: $I_{i+1}$ is an image of $I_i$ and the restriction of $I_{i}$ to its variables that do not appear in $I_{i+1}$ is acyclic. 

\newcommand{\untanglingqueryimagesmall}{
\node (A) at (1,0) {};
\node (B) at (.5,-.5) {};
\node (C) at (.5,-1) {};
\node (D) at (.1,-.75) {};
\node (E) at (1,-1.5) {};

\draw[->] (A) -- (B);
\draw[->] (B) -- (C);
\draw[->] (C) -- (D);
\draw[->] (D) -- (B); 
\draw[->] (C) -- (E);

\draw[] (.4,-.75) circle (11pt);
}
\newcommand{\untanglingqueryimage}{
\untanglingqueryimagesmall
\node (F) at (1.5,-1) {};

\draw[->] (F) -- (E);
}
\newcommand{\restsmall}{
\node (G) at (1.5,-.5) {};
\node (H) at (2,-.75) {};

\draw[->] (G) -- (F);
\draw[->] (F) -- (H);
\draw[->] (H) -- (G);
\draw[->] (A) -- (G);
}

\newcommand{\untanglingquery}{
\untanglingqueryimage
\restsmall
}

\begin{example}\label{example-untangling}
Consider the queries $q_6$ and $q_7$ depicted in \cref{figure-untangling-one}. They have the same acyclic core, namely the ternary relation and the three binary relations forming a triangle inside of it. The figure shows an untangling for $q_6$ and illustrates why there is no such untangling for $q_7$. Another example of a query that can be untangled is given in \cref{figure-untangling-three} in the appendix.

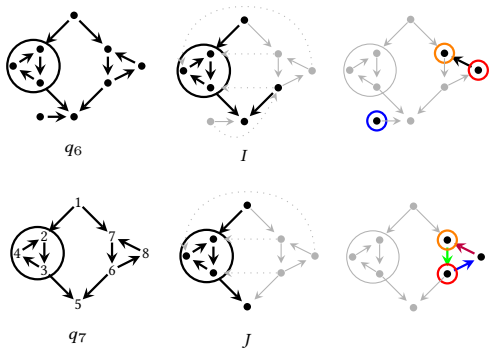
\begin{figure}[ht]
\begin{tikzpicture}[scale=.9]
\untanglingquery
\node (I) at (.5,-1.5) {};
\draw[->] (I) -- (E);
\node[nodevar] (Q) at (1,-2) {\var{$q_6$}};
\end{tikzpicture}
\hspace{.2cm}
 \begin{tikzpicture}[scale=.9]
\untanglingqueryimage
\node[nodevar] (Q) at (1,-2) {\var{$I$}};

\tikzset{every path/.style={color=black!30,shorten >=2pt}}
\node (I) at (.5,-1.5) {};
\draw[->] (I) -- (E);
\restsmall

\draw[dotted,->] plot [smooth,tension=1] coordinates {(I) (1,-1.7) (F)};
\draw[dotted,->] (F) -- (C);
\draw[dotted,->] (G) -- (B);
\draw[dotted,->] plot [smooth,tension=2] coordinates {(H) (1,.2) (D)};
\end{tikzpicture} \hspace{.2cm} \begin{tikzpicture}[scale=.9]
\node (G) at (1.5,-.5) {};
\node (H) at (2,-.75) {};
\node (I) at (.5,-1.5) {};

\draw[<-] (G) -- (H);

\draw[color=red] (H) circle (4pt);
\draw[color=orange] (G) circle (4pt);
\draw[color=blue] (I) circle (4pt);
\tikzset{every path/.style={color=black!30,shorten >=2pt}}
\untanglingqueryimage
\draw[->] (I) -- (E);
\draw[->] (G) -- (F);
\draw[->] (A) -- (G);
\draw[->] (F) -- (H);
\node[nodevar] (Q) at (1,-2) {};
\end{tikzpicture} 

\vspace{.4cm}

\begin{tikzpicture}[scale=.9]
\node[nodevar] (A) at (1,0) {\var{1}};
\node[nodevar] (B) at (.5,-.5) {\var{2}};
\node[nodevar] (C) at (.5,-1) {\var{3}};
\node[nodevar] (D) at (.1,-.75) {\var{4}};
\node[nodevar] (E) at (1,-1.5) {\var{5}};
\node[nodevar] (F) at (1.5,-1) {\var{6}};
\node[nodevar] (G) at (1.5,-.5) {\var{7}};
\node[nodevar] (H) at (2,-.75) {\var{8}};

\draw[->] (A) -- (B);
\draw[->] (B) -- (C);
\draw[->] (C) -- (D);
\draw[->] (D) -- (B); 
\draw[->] (C) -- (E);
\draw[] (.4,-.75) circle (11pt);

\draw[->] (F) -- (E);
\draw[->] (G) -- (F);
\draw[->] (F) -- (H);
\draw[->] (H) -- (G);
\draw[->] (A) -- (G);
\node[nodevar] (Q) at (1,-2) {\var{$q_7$}};
\end{tikzpicture}
\hspace{.2cm}
\begin{tikzpicture}[scale=.9]
\untanglingqueryimagesmall
\node[nodevar] (Q) at (1,-2) {\var{$J$}};

\tikzset{every path/.style={color=black!30,shorten >=2pt}}

\node (G) at (1.5,-.5) {};
\node (H) at (2,-.75) {};
\node (F) at (1.5,-1) {};

\draw[->] (F) -- (E);
\draw[->] (G) -- (F);
\draw[->] (F) -- (H);
\draw[->] (H) -- (G);
\draw[->] (A) -- (G);
\draw[dotted,->] (F) -- (C);
\draw[dotted,->] (G) -- (B);
\draw[dotted,->] plot [smooth,tension=2] coordinates {(H) (1,.2) (D)};
\end{tikzpicture} \hspace{.2cm} \begin{tikzpicture}[scale=.9]
\node (F) at (1.5,-1) {};
\node (G) at (1.5,-.5) {};
\node (H) at (2,-.75) {};

\draw[color=green,->] (G) -- (F);
\draw[color=blue,->] (F) -- (H);
\draw[color=purple,->] (H) -- (G);

\draw[color=orange] (G) circle (4pt);
\draw[color=red] (F) circle (4pt);
\tikzset{every path/.style={color=black!30,shorten >=2pt}}
\untanglingqueryimagesmall
\draw[->] (A) -- (G);
\draw[->] (F) -- (E);
\node[nodevar] (Q) at (1,-2) {};
\end{tikzpicture}
\caption{The query $q_6$ (the circle depicts a ternary relation) followed by an acyclic image $I$ and the result of the corresponding restriction, making the sequence $q_6,I$ an untangling of $q_6$. By \Cref{proposition-untangling}, $q_6$ can be enumerated with linear delay. The sequence $q_7,J$ is not an untangling of $q_7$ because the resulting query is cyclic. In fact, $q_7$ cannot be untangled and we show in \Cref{example-untangle-hard} that its enumeration is hard.
}
\label{figure-untangling-two}\label{figure-untangling-one}
\end{figure}

\begin{toappendix}
\begin{figure}[ht]
    \centering
\begin{tikzpicture}
\node (A) at (-1,.5) {};
\node (B) at (-1,-.5) {};
\node (C) at (-2,0) {};
\node (D) at (0,0) {};
\node (E) at (1,0.5) {};
\node (F) at (1,-0.5) {};
\node (G) at (2,0) {};

\draw[->] (A) -- (B);
\draw[->] (A) -- (C);
\draw[->] (C) -- (B);
\draw[->] (A) -- (D);
\draw[->] (D) -- (B);
\draw[->] (E) -- (F);
\draw[->] (E) -- (D);
\draw[->] (D) -- (F);
\draw[->] (E) -- (G);
\draw[->] (G) -- (F);
\draw[] (-1.4,0) circle (22pt);
\draw[dotted] (-.5,1) -- (-.5,-1);
\draw[dotted] (.5,1) -- (.5,-1);
\draw[dotted] (1.5,1) -- (1.5,-1);
\draw[dotted] (2.5,1) -- (2.5,-1);
\node[fill=none] at (-.75,1) {$I_3$};
\node[fill=none] at (.2,1) {$I_2$};
\node[fill=none] at (1.2,1) {$I_1$};
\node[fill=none] at (2.2,1) {$I_0$};
\end{tikzpicture}
    \caption{This query can be untangled, as witnessed by the sequence $I_0$, $I_1$, $I_2$ $I_3$, where the $I_i$ are the images corresponding to the part to the left of each dotted vertical line.}
    \label{figure-untangling-three}
\end{figure}
\end{toappendix}
\end{example}

\begin{theorem}\label{proposition-untangling}
Any full conjunctive query that has an untangling can be enumerated with linear delay (after linear preprocessing time).
\end{theorem}
\begin{proof}[Proof sketch]
Let $I_0,I_1,\cdots, I_k$ be an untangling of $q$.
The proof goes by induction on $k$. 
If $k=0$, then the query is acyclic and can be enumerated with linear delay by Theorem~\ref{dichotomy-sjf}. If $k>0$, consider the image $I_1$. Let $q'$ be the restriction of $q$ to its variables not in $I_1$. By construction, $q'$ is acyclic and can be enumerated with linear delay. By induction, $I_1$ can be enumerated with linear delay. We show that $q$ can then be enumerated with linear delay as well.

On an input database $D$, the enumeration of $q$ works as follows.
We start the enumeration for $I_1$. 
For each solution $\bar a$ to $I_1$, as witnessed by a mapping $\mu$ from the variables of $I_1$ to the elements of $D$, we compute in linear time a database $D'$ over the schema of $q'$ as follows.
By definition, each atom $R'(\bar x)$ of $q'$ is computed from an atom $R(\bar y)$ of $q$ by removing all variables in $I_1$. 
We filter $R(\vec{y})$ to only keep the tuples that ``agree'' with $\vec{a}$, and then we project the result to only keep the columns corresponding to $\vec{x}$. More formally,
let $\vec{u}$ be the variables of $\bar y$ occurring in $I_1$ and $\vec{v}$ those not occurring in $I_1$.
Then, $R'(D')$ consists of all the tuples $\nu(\vec{v})$ such that $\nu$ is a mapping from $\bar y$ to a tuple of $R$ with $\mu(\vec{u})=\nu(\vec{u})$. 

 We then start the enumeration for $q'$ on $D'$. Every solution to $q'$ can be combined with $\bar a$ to form a solution to $q$, and the algorithm outputs all solutions with no repetitions.
Since $I_1$ is an image, each $\bar a$ yields at least one solution to $q'$, and we get linear delay.
\end{proof}

\subsection{Necessary condition}\label{section-linear-hard}

We provide a condition identifying full queries that cannot be enumerated with linear delay, assuming \sparsehyperclique. We essentially reduce \sparsehyperclique to the computation of a constant number of solutions to the query in linear time. The construction is illustrated in the next example.

\begin{example}\label{example-untangle-hard}
Consider again the query $q_7$, depicted in \Cref{figure-untangling-two}. We show that we cannot produce $3$ solutions to $q_7$ in linear time unless we can test in $O(m)$ time whether a graph with $m$ edges contains a triangle (which is not possible assuming \sparsehyperclique).

Given an input graph $G$, we construct a database $D$ as follows.
For every edge $(a,b)$ of $G$ with $a<b$, we add to $D$ the facts $R(\pair{a,x_6},\pair{b,x_8})$, $R(\pair{b,x_8},\pair{a,x_7})$, $R(\pair{a,x_7},\pair{b,x_6})$,\\ $R(\pair{\bot,x_1},\pair{a,x_7})$, and $R(\pair{b,x_8},\pair{\bot,x_5})$. We also add to $D$ the facts $R(\bot,\bot)$, $R(\pair{\bot,x_1},\bot)$, $R(\bot,\pair{\bot,x_5})$, and $S(\bot,\bot,\bot)$. The number of facts of $D$ is linear in the number of edges of $G$, and this construction can be done in linear time in the size of $G$.

A case analysis on the tagging map, displayed in \Cref{query-q6}, shows that the data part of each solution to $q_7$ contains either the elements with $\bot$ or we can extract from it a triangle of $G$. As there are only a constant number of solutions of the first kind, any linear delay  algorithm will test the existence of a triangle of $G$ in linear time.

\begin{figure}[tb]
\centering
\begin{tikzpicture}[scale=1.1]
\node[nodevar] (A) at (1,0) {\var{$\bot$}};
\node[nodevar] (B) at (.5,-.5) {\var{$\bot$}};
\node[nodevar] (C) at (.5,-1) {\var{$\bot$}};
\node[nodevar] (D) at (.1,-.75) {\var{$\bot$}};
\node[nodevar] (E) at (1,-1.5) {\var{$\bot$}};
\node[nodevar] (F) at (1.5,-1) {\var{$\bot$}};
\node[nodevar] (G) at (1.5,-.5) {\var{$\bot$}};
\node[nodevar] (H) at (2,-.75) {\var{$\bot$}};

\draw[->] (A) -- (B);
\draw[->] (B) -- (C);
\draw[->] (C) -- (D);
\draw[->] (D) -- (B); 
\draw[->] (C) -- (E);
\draw[] (.4,-.75) circle (11pt);

\draw[->] (F) -- (E);
\draw[->] (G) -- (F);
\draw[->] (F) -- (H);
\draw[->] (H) -- (G);
\draw[->] (A) -- (G);
\end{tikzpicture}
\hspace{.1cm}
\begin{tikzpicture}[scale=1.1]
\tikzstyle{nodevar} = [fill=none, inner sep=0pt,rectangle,outer sep=0pt]
\node[nodevar] (A) at (1,0) {\var{$\pair{\bot,x_1}$}};
\node[nodevar] (B) at (.5,-.5) {\var{$\bot$}};
\node[nodevar] (C) at (.5,-1) {\var{$\bot$}};
\node[nodevar] (D) at (.1,-.75) {\var{$\bot$}};
\node[nodevar] (E) at (1,-1.5) {\var{$\pair{\bot,x_5}$}};
\node[nodevar] (F) at (1.5,-1) {\var{$\bot$}};
\node[nodevar] (G) at (1.5,-.5) {\var{$\bot$}};
\node[nodevar] (H) at (2,-.75) {\var{$\bot$}};

\draw[->] (A) -- (B);
\draw[->] (B) -- (C);
\draw[->] (C) -- (D);
\draw[->] (D) -- (B); 
\draw[->] (C) -- (E);
\draw[] (.4,-.75) circle (11pt);

\draw[->] (F) -- (E);
\draw[->] (G) -- (F);
\draw[->] (F) -- (H);
\draw[->] (H) -- (G);
\draw[->] (A) -- (G);
\end{tikzpicture}\hspace{.1cm}
\begin{tikzpicture}[scale=1.1]
\tikzstyle{nodevar} = [fill=none, inner sep=0pt,rectangle,outer sep=0pt]
\node[nodevar] (A) at (1,0) {\var{$\pair{\bot,x_1}$}};
\node[nodevar] (B) at (.5,-.5) {\var{$\bot$}};
\node[nodevar] (C) at (.5,-1) {\var{$\bot$}};
\node[nodevar] (D) at (.1,-.75) {\var{$\bot$}};
\node[nodevar] (E) at (1,-1.5) {\var{$\pair{\bot,x_5}$}};
\node[nodevar] (F) at (1.5,-1) {\var{$\pair{b,x_6}$}};
\node[nodevar] (G) at (1.5,-.5) {\var{$\pair{a,x_7}$}};
\node[nodevar] (H) at (2.5,-.75) {\var{$\pair{c,x_8}$}};

\draw[->,outer sep=0pt] (A) -- (B);
\draw[->] (B) -- (C);
\draw[->] (C) -- (D);
\draw[->] (D) -- (B); 
\draw[->] (C) -- (E);
\draw[] (.4,-.75) circle (11pt);

\draw[->,outer sep=0pt] (F) -- (E);
\draw[->,inner sep=-2pt] (G) -- (F);
\draw[->] (F) -- (H);
\draw[->] (H) -- (G);
\draw[->] (A) -- (G);
\end{tikzpicture}
\caption{Illustrations of solutions to $q_7$ obtained by the core tagging map, the tagging that maps the right side to the left, and the identity tagging map.}\label{query-q6}
\end{figure}
\end{example}

\Cref{theorem-necessary-linear} builds upon the idea underlined in Example~\ref{example-untangle-hard}, and provides a necessary condition for linear delay tractability.
An endomorphism of a query is said to be \emph{idempotent} if it is the identity on its image. 
As an example, in \Cref{figure-untangling-one} the endomorphism witnessing $J$ is idempotent while the one for $I$ is not.

\begin{theoremrep}\label{theorem-necessary-linear}
Let $q$ be a full conjunctive query. If $q$ contains a set $S$ of variables such that the restriction $\projectedq{q}{S}$ is cyclic, and all its endomorphisms are idempotent and have an image containing either all the variables in $S$ or none of them, then $q$ cannot be enumerated with linear delay (and linear preprocessing), assuming \sparsehyperclique.
\end{theoremrep}
\begin{appendixproof}
According to \Cref{dichotomy-sjf}, the testing problem for $\projectedq{q}{S}$ cannot be solved in linear time since it is cyclic and self-join free, assuming \sparsehyperclique. In particular, it cannot be enumerated in linear delay. According to the following \Cref{proposition-linear-hard-ter}, this implies the same for $q$.
\begin{lemma}\label{proposition-linear-hard-ter}
 Let $q$ be a full conjunctive query whose endomorphisms are all idempotent, and let $S$ be a set of variables of $q$ such that every image of $q$ contains either all of $S$ or none of them. Then, $q$ is at least as hard for enumeration as $\projectedq{q}{S}$. 
\end{lemma}
\begin{proof}[Proof of \Cref{proposition-linear-hard-ter}]
Let $D'$ be a database for $\projectedq{q}{S}$. We construct a database $D$ for $q$ as follows.
Consider an atom $R(\vec{y})$ of $q$, and let $R'(\vec{x})$ be its restriction in $\projectedq{q}{S}$. For every fact $R'(\vec{a})$, we construct the tuple $\vec{b}$ from $\vec{a}$ by inserting $\bot$ in all positions of $\vec{y}$ that do not contain variables of $S$.
We then insert to $D$ the fact $R(\bar c)$ with $c_i=\pair{b_i,y_i}$.
Notice that, by construction, the elements of $D$ are pairs of the form $\pair{a,x}$ such that $a$ is either $\bot$ or an element of $D'$ and $x$ is a variable of $q$. Moreover, whenever $x \in S$ then $a$ is an element of $D'$.

Consider now an answer $\bar a$ of $q$ on $D$. Let $\mu$ be the mapping witnessing the fact that $\bar a$ is a solution. As in the tagging technique, we perform a case analysis depending on the endomorphism $\nu$ of $q$ induced by $\mu$. If the image of $\nu$ does not intersect with $S$, then the data part of $\mu$ contains only the elements $\bot$ and, as there is only a constant number of such solutions, we can safely ignore them. Otherwise, by our assumption, the image of $\nu$ must contain $S$, and $\nu$ is the identity on $S$.
This implies that for any variable $x\in S$, $\mu(x)$ is of the form $\pair{b,x}$ where $b$ is an element of $D'$. We claim that the mapping $\mu'$ defined by $\mu'(x)=b$ whenever $x\in S$ and $\mu(x)=(b,x)$ witnesses a solution of $\projectedq{q}{S}$ on $D'$.
For all $R'(\vec{x})$ in $\projectedq{q}{s}$, consider $R(\vec{y})$. Let $\vec{u}$ be the variables of $y$ occurring in $S$. Since $\mu(\vec{y})\in R$ and $\nu$ is idempotent, we have that $\mu(\vec{u})$ is a tuple of $R'$ tagged by $\vec{x}$. Hence, our construction ``filters enough'', and the data part gives a solution of $\projectedq{q}{S}$. Notice also that all solutions to $\projectedq{q}{S}$ can be obtained this way. 
Hence, the enumeration of $q$ will enumerate all the solutions to $\projectedq{q}{S}$.
\end{proof}
\end{appendixproof}
\begin{proofsketch}
We show that the enumeration for $q$ is at least as hard as the enumeration for $\projectedq{q}{S}$, using a construction that assigns the constant $\bot$ to the extra variables and a case analysis according to the endomorphism producing each answer. The idempotency and the fact that each image that contains a variable of $S$ contains all variables of $S$ guarantee that each answer that contains a non-constant value corresponds to an answer to $\projectedq{q}{S}$.
\end{proofsketch}

\subsection{Other cases}
The conditions from \Cref{section-linear-easy} and \Cref{section-linear-hard} do not cover all queries.
We conclude this section with an example-driven illustration of the difficulty in achieving a full classification for full queries. We first give one example of a query that can be enumerated with linear delay but cannot be untangled. We next give an example of a query that cannot be enumerated with linear delay but does not fulfill the condition of \Cref{theorem-necessary-linear}. Finally, we specify a third query for which the complexity we do not yet know.

\newcommand{\hardcore}{
\node[nodevar] (A) at (0,0) {\var{1}};
\node[nodevar] (G) at (-.5,-.5) {\var{2}};
\node[nodevar] (H) at (-.8,.4) {\var{3}};

\draw[->] (A) -- (G);
\draw[->] (G) -- (H);
\draw[->] (H) -- (A);

\draw[] (-.5,0) circle (20pt);
}
\newcommand{\hardtopimage}{ 
\hardcore
\node[nodevar] (B) at (1,.5) {\var{4}};
\node[nodevar] (D) at (2,0) {\var{5}};
\node[nodevar] (E) at (1.5,1) {\var{6}};

\draw[->] (A) -- (B);
\draw[->] (B) -- (D);
\draw[->] (D) -- (E);
\draw[->] (E) -- (B);
}

\newcommand{\hardbottomimage}{ 
\hardcore
\node[nodevar] (C) at (1,-.5) {\var{7}};
\node[nodevar] (D) at (2,0) {\var{5}};
\node[nodevar] (F) at (1.5,-1) {\var{8}};

\draw[->] (A) -- (C);
\draw[->] (C) -- (D);
\draw[->] (D) -- (F);
\draw[->] (F) -- (C);
}

\begin{figure}[htb]
\centering
\begin{tikzpicture}[scale=.7]
\hardtopimage
\node[nodevar] (C) at (1,-.5) {\var{7}};
\node[nodevar] (F) at (1.5,-1) {\var{8}};

\draw[->] (A) -- (C);
\draw[->] (C) -- (D);
\draw[->] (D) -- (F);
\draw[->] (F) -- (C);

\node[nodevar] (Q) at (0,-1.8) {\var{$q_8$}};
\end{tikzpicture}
\hspace{0.5cm}
\begin{tikzpicture}[scale=.7]
\node[nodevar] (A) at (0,0) {\var{2}};
\node[nodevar] (C) at (0,2) {\var{3}};
\node[nodevar] (D) at (1,1) {\var{1}};
\node[nodevar] (E) at (-1,1) {\var{4}};

\draw[->] (A) -- (C);
\draw[->] (C) -- (D);
\draw[->] (D) -- (A);
\draw[->] (E) -- (A);
\draw[->] (E) -- (C);

\draw[->] (D) to[in=-40,out=40,loop,min distance=1cm] (D);
\draw[->] (E) to[in=-120,out=120,loop,min distance=1cm] (E);
\node[nodevar] (Q) at (0,-.8) {\var{$q_9$}};
\end{tikzpicture}
\hspace{0.5cm}
\begin{tikzpicture}[scale=.7]
\node (A) at (0,0) {};
\node (B) at (0,1) {};
\node (C) at (0,2) {};
\node (D) at (1,1) {};
\node (E) at (-1,1) {};

\draw[->] (A) -- (B);
\draw[->] (B) -- (C);
\draw[->] (C) -- (D);
\draw[->] (D) -- (A);
\draw[->] (E) -- (A);
\draw[->] (E) -- (C);

\draw[->] (D) to[in=-40,out=40,loop,min distance=1cm] (D);
\draw[->] (E) to[in=-120,out=120,loop,min distance=1cm] (E);
\node[nodevar] (Q) at (0,-.8) {\var{$q_{10}$}};
\end{tikzpicture}
\caption{The hard query $q_8$ of \Cref{example:hard-query}, the easy query $q_9$ of \Cref{example-easy-twotriangles}, and the unresolved query $q_{10}$ of \Cref{example-open}.}\label{figure-beyond-nested}
\end{figure}

\begin{example}\label{example:hard-query}
Consider the query $q_8$ of \Cref{figure-beyond-nested}. The core of the query is the ternary atom and the triangle within it and is therefore acyclic.
The query cannot be untangled because the core is the only acyclic image and the result of the associated restriction yields a query with a cyclic core.
The reader can also verify that the condition of \cref{theorem-necessary-linear} is not met.
Nevertheless, $q_8$ cannot be enumerated with linear delay  unless we can find a triangle in a graph in time linear in the number of its edges, a special case of \sparsehyperclique. The proof idea is that we encode the triangles of the graph into both query triangles that are not covered by a ternary relation and assign constants to all other variables. Except for one solution, all solutions over this encoding identify triangles in the input graph, though they do not all originate in the same query triangle. The details can be found in the appendix.
\begin{toappendix}
{\bf Missing details in \cref{example:hard-query}.}
The condition of \cref{theorem-necessary-linear} is not met because any set $S$ of variables containing a cycle must contain the variable $x_5$; if it also contains $x_7$ and $x_8$ then the image $x_1,\cdots,x_6$ intersects with $S$ in $x_5$ but does not contain $x_7$. The case where $S$ contains $x_4$ and $x_6$ is symmetric.

The query $q_8$ cannot be enumerated with linear delay  unless we can find a triangle in a graph in time linear in the number of its edges, a special case of \sparsehyperclique.
To show this, We use the tagging technique as follows.
Given an input graph $G$, we construct in linear time the following database $D$. To each edge $(a,b)$ with $a<b$, we add to $D$ the facts $R(\pair{a,x_4},\pair{b,x_5})$, $R(\pair{a,x_5},\pair{b,x_6})$, $R(\pair{b,x_6},\pair{a,x_4})$, $R(\pair{a,x_5},\pair{b,x_8})$, $R(\pair{b,x_8},\pair{a,x_7})$, $R(\pair{a,x_7},\pair{b,x_5})$. We also add to $D$ the facts $R(\pair{\bot,x_1},\pair{a,x_4})$ and $R(\pair{\bot,x_1},\pair{a,x_7})$. Finally we also add the facts $R(\pair{\bot,x_1},\pair{\bot,x_2})$, $R(\pair{\bot,x_2},\pair{\bot,x_3})$, $R(\pair{\bot,x_3},\pair{\bot,x_1})$ and $S(\pair{\bot,x_1},\pair{\bot,x_2},\pair{\bot,x_3})$. Consider now a solution to $q$ over $D$. Due to the tagging technique, the solution induces an endomorphism of $q$ (the tagging map). Let $I$ be the corresponding image. If $I$ is any image but the core, the solution to $q$ induces a triangle of $G$. If $I$ is the core, then the data part of the solution only contains $\bot$, and there is only one such solution. Hence, if we can enumerate the solutions to the query with linear delay and linear preprocessing, then we can detect in linear time the presence of a triangle in $G$.

 \begin{figure}[ht]
 \centering
 \begin{tikzpicture}[scale=1]
 \hardtopimage
 \node[nodevar] (C) at (1,-.5) {\var{7}};
 \node[nodevar] (F) at (1.5,-1) {\var{8}};

 \draw[->] (A) -- (C);
 \draw[->] (C) -- (D);
 \draw[->] (D) -- (F);
 \draw[->] (F) -- (C);

 \end{tikzpicture}
 \hspace{1.5cm}
 \begin{tikzpicture}[scale=1]
 \hardtopimage
 \tikzset{every path/.style={color=black!30,shorten >=2pt}}
\node[nodevar] (C) at (1,-.5) {\var{7}};
 \node[nodevar] (F) at (1.5,-1) {\var{8}};

 \draw[->] (A) -- (C);
 \draw[->] (C) -- (D);
 \draw[->] (D) -- (F);
 \draw[->] (F) -- (C);
\draw[dotted,->] plot [smooth,tension=1] coordinates {(F) (2.5,0) (E)};
\draw[dotted,->] (C) -- (B);

 \end{tikzpicture}
 \hspace{1.5cm}
 \begin{tikzpicture}[scale=1]
 \hardbottomimage
  \tikzset{every path/.style={color=black!30,shorten >=2pt}}
\node[nodevar] (B) at (1,.5) {\var{4}};
 \node[nodevar] (E) at (1.5,1) {\var{6}};

 \draw[->] (A) -- (B);
 \draw[->] (B) -- (D);
 \draw[->] (D) -- (E);
 \draw[->] (E) -- (B);
\draw[dotted,->] plot [smooth,tension=1] coordinates {(E) (2.5,0) (F)};
\draw[dotted,->] (B) -- (C);

 \end{tikzpicture}
 \caption{From left to right: query $q_8$, its ``top'' image, and its ``bottom'' image.}\label{figure-hard-query}
 \end{figure}
 
\end{toappendix}
\end{example}

\begin{example}\label{example-easy-twotriangles} Consider the query $q_9$ depicted in \Cref{figure-beyond-nested}. Its core is a self-loop and is acyclic. One can verify that this query cannot be untangled. However, its solutions can be efficiently enumerated as follows. For each edge $(b,c)$ of the database, we maintain two lists: the list of self-loops $a$ such that $(a,b)$ and $(c,a)$ are edges, and a similar list of self-loops where $(a,b)$ and $(a,c)$ are edges.
The algorithm goes through all self-loops $a$, and for each such self-loop, we consider all edges $(b,c)$. If the triple $(a,b,c)$ satisfies any of the two patterns, we add $a$ to the corresponding list for $(b,c)$. In addition, if $(a,b,c)$ satisfies the first pattern, then for each $a'$ in the second list for $(b,c)$, we output the solution $(a,b,c,a')$. Similarly, if $(a,b,c)$ satisfies the second pattern, then for each $a'$ in the first list for $(b,c)$, we output the solution $(a',b,c,a)$ (unless $a'=b=c=a$ for avoiding duplicates). It can be easily verified that all solutions are printed and that no solution is printed twice. Moreover, as every self-loop $a$ results in a solution $(a,a,a,a)$, the algorithm exhibits linear delay. It is worth noting that, as before, the memory this algorithm uses may be linear in the output size (which may be larger than linear in the database size).
\end{example}

\begin{example}\label{example-open}
Finally, consider the query $q_{10}$ of \Cref{figure-beyond-nested}. It cannot be untangled. The ideas used in \cref{example-easy-twotriangles} can no longer be used as the middle part is now a path of length two and there may be more than a linear number of those. The idea from \Cref{example:hard-query} does not work either because similarly encoding triangles yields a linear number of non-triangle solutions. Whether $q_{10}$ can be enumerated with linear delay remains open. 
\end{example}

\section{Full Conjunctive Queries and Constant Delay}\label{section-constant}

In this section, we aim to distinguish the queries that can be enumerated with constant delay after a linear time preprocessing from those where this is not possible.
As in the previous section, all our queries will be full cyclic queries with an acyclic core.

\subsection{Mirror queries}

We start by showing that some cyclic queries can be solved with constant delay following linear preprocessing due to self-joins. This happens when half of the query can be seen as a reflection of the other half, as demonstrated in the following example.

\begin{example}\label{example-mirror}
The cyclic query $q_1$ of Figure~\ref{figure-diamonds} can be enumerated with constant delay as we describe next. First notice that the core $p$ of $q_1$, depicted by \begin{tikzpicture}
\node[nodevar] (A) at (0,0) {\var{1}};\node[nodevar] (B) at (.7,0) {\var{2}};\node[nodevar] (C) at (1.4,0) {\var{3}};\draw[->] (A) -- (B);\draw[->] (B) -- (C);
\end{tikzpicture}, is acyclic and free-connex, and so by Theorem~\ref{dichotomy-sjf}, $p$ can be enumerated with constant delay after linear preprocessing time. We will maintain a table, initialized as empty, that on entry a pair $(a,b)$ returns a list of elements $c$ such that $(a,c,b)$ is a solution to $p$.
The enumeration of $q_1$ works as follows. We perform the preprocessing necessary for $p$, start its enumeration, and for every solution $(a,c,b)$ we do the following. First, we output the solution $(a,c,b,c)$. Then, for every $c'$ that is given by the table for the entry $(a,b)$, we output the solutions $(a,c,b,c')$ and $(a,c',b,c)$. Finally, we add $c$ to the table entry $(a,b)$. We get that the delay is constant, and every result is printed once.
%Notice also that this algorithm requires the table and therefore uses a memory that may not be linear in the input size during the enumeration process.
\end{example}

\cref{example-mirror} puts forward the following sufficient condition.
We call a query a \emph{mirror} if it has an acyclic image $I$ such that the remaining atoms form a query isomorphic to $I$, where the isomorphism is the identity on the variables shared between the image and the remaining atoms.

\begin{propositionrep}\label{proposition-mirror}
Any full mirror conjunctive query can be enumerated with constant delay after a linear preprocessing time.
\end{propositionrep}
\begin{proof}
We essentially do as in \cref{example-mirror}.
Let $I$ be the acyclic image witnessing the mirror, and let $\bar x$ be the shared variables (that is, variables that appear in $I$ and also in atoms of the query that are not in $I$). 
Assume the free variables are ordered such that $\bar x$ appear first, then the other image variables, and then the remaining variables.
We will maintain a table that upon entry $\bar a$ returns a list of tuples $\bar b$ such that $\bar a\bar b$ forms a solution to $I$.
As $I$ is full and acyclic, it can be enumerated with constant delay after a linear preprocessing time. We perform the preprocessing of $I$ and start its enumeration. For every solution $\bar a\bar b$ to $I$, we do the following. First, we output the solution $\bar a\bar b\bar b$. Then, for every $b'$ given by the table for $\bar a$, we output the solutions $\bar a\bar b\bar b'$ and $\bar a\bar b'\bar b$. Finally, we add $\bar b$ to the table with entry $\bar a$. Notice that the delay is constant, and all solutions are printed with no repetition.
\end{proof}
As in \cref{example-mirror}, in the proof of \cref{proposition-mirror} we use during the enumeration  a memory linear in the output size (which may not be linear in the input size).

\subsection{The effect of unary atoms}

In contrast to the self-join-free case, we show that unary atoms can affect query complexity. Intuitively, when adding a unary atom to one of the sides of a mirror query, the solutions to this side no longer form all possible solutions to the other side, and so the algorithm we proposed for mirror queries fails.
The following example differs from the previous one only by the addition of a unary atom.

\begin{example}\label{example-unary-constant}
Consider again the query $q_2$ depicted in Figure~\ref{figure-diamonds}. We have seen in Example~\ref{example-acycliccore-linear} that it can be enumerated with linear delay. It turns out that it cannot be enumerated with linear preprocessing and constant delay unless we can test in linear time whether a graph contains a triangle, contradicting \sparsehyperclique. Given a graph $G$ with $m$ edges, we construct a database $D$ as follows. For every edge $(a,b)$ of $G$ with $a<b$, we add to $D$ the facts $R(\pair{a,x_1},\pair{b,x_2})$, $R(\pair{b,x_2},\pair{b,x_3})$, $R(\pair{a,x_1},\pair{b,x_4})$, and $R(\pair{b,x_4},\pair{a,x_3})$, together with the fact $\text{Red}(\pair{b,x_2})$. By considering all endomorphisms of $q_2$, as we explained when introducing the tagging technique, we get that a solution in $q(D)$ is either obtained by the identity automorphism and so it is of the form $(\pair{a,x_1},\pair{b,x_2},\pair{b,x_3},\pair{c,x_4})$ for a triangle $(a,b,c)$ in $G$, or it is obtained by the endomorphism that maps $x_4$ to $x_2$ and so it is of the form $(\pair{a,x_1},\pair{b,x_2},\pair{b,x_3},\pair{b,x_2})$ for an edge $(a,b)$ in $G$.
As there are at most $m$ solutions of the latter kind, any constant delay enumeration of $q$ in $D$ would be able to test in $O(m)$ whether $G$ has a triangle.
\end{example}

This added difficulty is caused by any structure that breaks the symmetry between the two sides. As an example, we could replace the unary atom in $q_2$ by a directed path of length $2$ starting in $x_2$ and still obtain the hardness with a similar proof.

\subsection{The effect of ``spikes''}

While introducing unary atoms can make queries harder, we will now see that introducing ``dangling'' binary atoms may make queries easier.  
Consider the query $q_5$ of \Cref{figure-blowfish} together with the three queries depicted in Figure~\ref{figure-blowfish-spikes}.
Notice that all four agree on the central loop and have the same acyclic core \begin{tikzpicture}
\core
\end{tikzpicture}. They can be untangled and therefore enumerated with linear delay due to \Cref{proposition-untangling}.
We prove in the sequel that $q_{11}$ and $q_{12}$ can even be enumerated with constant delay, while $q_5$ and $q_{13}$ cannot without a major computational breakthrough.

\begin{figure}[htb]
\centering
\begin{tikzpicture}[scale=.8]
\tikzset{mynode/.style={color=white}}
\blowfish
\node[mynodeA] (S4) at (1.6,-1) {};
\node[mynodeA] (S6) at (.5,-2.5) {};

\draw[->] (E) -- (S6);
\draw[<-] (C) -- (S4);

\node[nodevar] (Q) at (0,-3) {$q_{11}$};

\end{tikzpicture}
\hspace{.5cm}
\begin{tikzpicture}[scale=.8]
\tikzset{mynodeAA/.style={color=white}}
\blowfisheasy
\draw[->] (A) -- (S1);
\draw[<-] (CC) -- (S2);
\draw[->] (B) -- (S3);
\draw[<-] (DD) -- (S5);
\draw[->] (E) -- (S6);
\draw[->] (E) -- (S7);

\node[nodevar] (Q) at (0,-3) {$q_{12}$};

\end{tikzpicture}\hspace{.1cm}
\begin{tikzpicture}[scale=.8]
\tikzset{mynodeAA/.style={color=white}}
\blowfisheasy
\draw[->] (A) -- (S1);
\draw[<-] (CC) -- (S2);
\draw[->] (B) -- (S3);
\draw[->] (DD) -- (S5);
\draw[->] (E) -- (S6);
\draw[->] (E) -- (S7);

\node[nodevar] (Q) at (0,-3) {$q_{13}$};

\end{tikzpicture}

\caption{Queries similar to $q_5$ with a different status regarding constant delay enumeration.
}\label{figure-blowfish-spikes}
\end{figure}

Removing the unary atom from $q_5$ results in a mirror query that can be enumerated with constant delay according to \Cref{proposition-mirror}. This unary atom renders $q_5$ difficult, and the proof (in the appendix) is similar to that of \Cref{example-unary-constant}.
\begin{toappendix}

{\bf Missing details for the query $q_5$}.
The query $q_5$ of \cref{figure-blowfish-spikes} cannot be enumerated with constant delay assuming the hardness of triangle detection as conjectured in \sparsehyperclique. Let $G$ be a graph. We construct the following database $D$. For every node $a$ of $G$, we add the following facts to $D$: $R(\pair{a,x_1},\pair{a,x_2})$, $R(\pair{a,x_2},\pair{a,x_3})$, $R(\pair{a,x_4},\pair{a,x_3})$, $R(\pair{a,x_5},\pair{a,x_4})$,  $R(\pair{a,x_1},\pair{a,x_8})$, and $Red(\pair{a,x_2})$. For every edge $(a,b)$ of $G$ with $a<b$, we add the following facts to $D$: $R(\pair{a,x_5},\pair{b,x_6})$, $R(\pair{a,x_6},\pair{b,x_7})$, $R(\pair{a,x_8},\pair{b,x_7})$. Consider now a solution in $q_5(D)$. It induces an endomorphism of $q_5$ as explained in \Cref{section-trick}. We do a case analysis depending on the image $I$ of this endomorphism. If $I$ is the core,  then the data part of the solution involves only one node $a$ of $G$. The same holds if $I$ involves only the left part of $q_5$:  the variables from $x_1$ to $x_5$. 
If $I$ involves the top part of $q_5$, the variables $x_1,x_2,x_3,x_8,x_7$, then the data part of the solution corresponds to an edge of $G$. The data part of all the remaining images induces a triangle in $G$. Hence, the data part of all the solutions in $q_5(D)$  contain a triangle of $G$ except for a number of solutions linear in the size of $G$. Hence, a constant delay enumeration algorithm for $q_5$ with linear preprocessing would induce a linear test of whether $G$ contains a triangle.

\vspace*{0.5\baselineskip}
The reader may verify that the encoding used to prove the hardness of $q_5$ does not work for $q_{11}$ as the spikes make bigger images. With this encoding, there would be one solution per path of distance two in $G$, and so the triangles might be detected only after quadratic time, which does not contradict \sparsehyperclique.

\end{toappendix}
The spikes of $q_{11}$ make it so that all cycle edges appear in an acyclic image. This can be used to devise an efficient algorithm as we explain next.

\begin{figure}[tb]
    \centering
    \begin{tikzpicture}[scale=.8]
    \tikzset{mynode/.style={color=white}}
\blowfish
\node[nodevar] (S4) at (1.6,-1) {\var{10}};
\node[nodevar] (S6) at (.5,-2.5) {\var{9}};

\draw[->] (E) -- (S6);
\draw[<-] (C) -- (S4);
\end{tikzpicture}
\hspace{0.4cm}
\begin{tikzpicture}[scale=.8]
\node[nodevar] (A) at (0,0) {\var{1}};
\node[nodevar] (D) at (.5,-1.5) {\var{6}};
\node[nodevar] (BB) at (-.5,-.5) {\var{2}};
\node[nodevar] (CC) at (-1,-1) {\var{3}};
\node[nodevar] (E) at (0,-2) {\var{5}};
\node[nodevar] (DD) at (-.5,-1.5) {\var{4}};

\draw[->] (E) -- (DD);
\draw[->] (DD) -- (CC);
\draw[->] (E) -- (D);
\draw[->] (A) -- (BB);
\draw[->] (BB) -- (CC);
\draw[color=red] (BB) circle (5pt);
\tikzset{every path/.style={color=black!30,shorten >=2pt}}
\node[nodevar] (S4) at (1.6,-1) {\var{10}};
\node[nodevar] (S6) at (.5,-2.5) {\var{9}};
\node[nodevar] (B) at (.5,-.5) {\var{8}};
\node[nodevar] (C) at (1,-1) {\var{7}};

\draw[->] (D) -- (C);
\draw[->] (A) -- (B);
\draw[->] (B) -- (C);
\draw[->] (E) -- (S6);
\draw[<-] (C) -- (S4);

\tikzset{every path/.style={color=black!30,shorten >=2pt}}
\draw[dotted,->] (B) -- (BB);
\draw[dotted,->] (C) -- (CC);
\draw[dotted,->] (D) -- (DD);
\draw[dotted,->] plot [smooth,tension=1] coordinates {(S4) (0,.5) (BB)};
\draw[dotted,->] (S6) -- (D);
\end{tikzpicture}
%\hspace{0.1cm}
\begin{tikzpicture}[scale=.8]
\node[nodevar] (A) at (0,0) {\var{1}};
\node[nodevar] (B) at (.5,-.5) {\var{8}};
\node[nodevar] (C) at (1,-1) {\var{7}};
\node[nodevar] (D) at (.5,-1.5) {\var{6}};
\node[nodevar] (BB) at (-.5,-.5) {\var{2}};
\node[nodevar] (CC) at (-1,-1) {\var{3}};

\draw[->] (A) -- (B);
\draw[->] (B) -- (C);
\draw[->] (D) -- (C);
\draw[->] (A) -- (BB);
\draw[->] (BB) -- (CC);
\draw[color=red] (BB) circle (5pt);
\tikzset{every path/.style={color=black!30,shorten >=2pt}}
\node[nodevar] (DD) at (-.5,-1.5) {\var{4}};
\node[nodevar] (S4) at (1.6,-1) {\var{10}};
\node[nodevar] (S6) at (.5,-2.5) {\var{9}};
\node[nodevar] (E) at (0,-2) {\var{5}};

\draw[->] (E) -- (D);

\draw[->] (E) -- (DD);
\draw[->] (DD) -- (CC);
\draw[->] (E) -- (S6);
\draw[<-] (C) -- (S4);
\draw[dotted,->] (DD) -- (BB);
\draw[dotted,->] (E) -- (A);
\draw[dotted,->] (D) -- (B);
\draw[dotted,->] plot [smooth,tension=1] coordinates {(S4) (1.2,-1.3) (D)};
\draw[dotted,->] plot [smooth,tension=2] coordinates {(S6) (-1.5,-1) (BB)};

\end{tikzpicture} 
    \caption{The query $q_{11}$, its left image, and its top image.}
    \label{figure-qtwo}
\end{figure}
\begin{example}
Consider $q_{11}$ and its images of \cref{figure-qtwo}. The top and left images can be enumerated with linear preprocessing and constant delay as they are acyclic and free-connex. We start by enumerating the solutions of the left image, {and maintain a lookup table as follows}. To any solution $(a,b,c,d,e,f)$ of the left image, we add the pair $(d,e)$ to the table at the entry $(a,b,c,f)$. We output the solution $(a,b,c,d,e,d,c,b,f,b)$ induced by this image. Once we are done with the left image, we start the enumeration of the top image. To any solution $(a,b,c,d,e,f)$ of the top image, we output the solution $(a,b,c,b,a,f,e,f,b,d)$ induced by this image. Moreover, for every pair $(u,v)$ in the table for the entry $(a,b,c,d)$, the tuple $(a,b,c,u,v,d,e,f)$ is a solution for the main loop. We go through all edges outgoing from $v$ and all edges incoming to $e$ to get a solution to $q_{11}$ in constant delay. Altogether, we get a constant delay algorithm that outputs all solutions to $q_{11}$, possibly $3$ times. We remove the duplicates using the Cheater's Lemma.
\end{example}

We show next that $q_{12}$ can also be efficiently solved even though it does not have acyclic images covering all edges of the main loop. 

\begin{figure}[tb]
    \centering
\begin{tikzpicture}[scale=.8]
\node[nodevar] (A) at (0,0) {\var{1}};
\node[nodevar] (B) at (.5,-.5) {\var{8}};
\node[nodevar] (BB) at (-.5,-.5) {\var{2}};
\node[nodevar] (CC) at (-1,-1) {\var{3}};
\node[nodevar] (DD) at (-.5,-1.5) {\var{4}};

\draw[->] (A) -- (B);
\draw[->] (A) -- (BB);
\draw[->] (BB) -- (CC);
\draw[->] (DD) -- (CC);
\draw[color=red] (BB) circle (5pt);
\tikzset{every path/.style={color=black!30,shorten >=2pt}}
\node[nodevar] (C) at (1,-1) {\var{7}};
\node[nodevar] (D) at (.5,-1.5) {\var{6}};
\node[nodevar] (E) at (0,-2) {\var{5}};
\node[nodevar] (S1) at (0,.6) {\var{9}};
\node[nodevar] (S2) at (-1.6,-1) {\var{10}};
\node[nodevar] (S3) at (1,0) {\var{14}};
\node[nodevar] (S5) at (-1,-2) {\var{11}};
\node[nodevar] (S6) at (.5,-2.5) {\var{13}};
\node[nodevar] (S7) at (-.5,-2.5) {\var{12}};

\draw[->] (E) -- (D);
\draw[->] (B) -- (C);
\draw[->] (D) -- (C);
\draw[->] (E) -- (DD);
\draw[->] (E) -- (S6);
\draw[->] (A) -- (S1);
\draw[->] (S2) -- (CC);
\draw[->] (B) -- (S3);
\draw[->] (S5) -- (DD);
\draw[->] (E) -- (S7);

\tikzset{every path/.style={color=black!30,shorten >=2pt}}
%\draw[dotted,->] (B) -- (BB);
%\draw[dotted,->] (C) -- (CC);
%\draw[dotted,->] (DD) -- (BB);
%\draw[dotted,->] (E) -- (A);
%\draw[dotted,->] (D) -- (BB);

%\draw[dotted,->] plot [smooth,tension=2] coordinates {(S6) (1.8,-1) (B)};
\draw[dotted,->] plot [smooth,tension=1] coordinates {(S1) (.4,0) (B)};
%\draw[dotted,->] plot [smooth,tension=1] coordinates {(S3) (0,1) (CC)};
\draw[dotted,->] plot [smooth,tension=1] coordinates {(S2) (-1,-1.4) (DD)};
%\draw[dotted,->] plot [smooth,tension=1] coordinates {(S5) (-2,-1) (A)};
%\draw[dotted,->] plot [smooth,tension=1] coordinates {(S7) (0,-1) (BB)};

\end{tikzpicture}
\hspace{.5cm}
\begin{tikzpicture}[scale=.8]
\node[nodevar] (A) at (0,0) {\var{1}};
\node[nodevar] (B) at (.5,-.5) {\var{8}};
\node[nodevar] (D) at (.5,-1.5) {\var{6}};
\node[nodevar] (E) at (0,-2) {\var{5}};
\node[nodevar] (BB) at (-.5,-.5) {\var{2}};
\node[nodevar] (CC) at (-1,-1) {\var{3}};
\node[nodevar] (DD) at (-.5,-1.5) {\var{4}};
\node[nodevar] (S1) at (0,.6) {\var{9}};
\node[nodevar] (S2) at (-1.6,-1) {\var{10}};
\node[nodevar] (S5) at (-1,-2) {\var{11}};
\node[nodevar] (S7) at (-.5,-2.5) {\var{12}};

\draw[->] (A) -- (S1);
\draw[->] (S2) -- (CC);
\draw[->] (S5) -- (DD);
\draw[->] (E) -- (S7);

\draw[->] (A) -- (BB);
\draw[->] (BB) -- (CC);
\draw[->] (DD) -- (CC);
\draw[->] (E) -- (DD);
\draw[->] (E) -- (D);
\draw[color=red] (BB) circle (5pt);

\tikzset{every path/.style={color=black!30,shorten >=2pt}}
\node[nodevar] (C) at (1,-1) {\var{7}};
\node[nodevar] (S3) at (1,0) {\var{14}};
\node[nodevar] (S6) at (.5,-2.5) {\var{13}};

\draw[->] (A) -- (B);
\draw[->] (B) -- (C);
\draw[->] (D) -- (C);
\draw[->] (B) -- (S3);
\draw[->] (E) -- (S6);

\tikzset{every path/.style={color=black!25,shorten >=2pt}}
%\draw[dotted,->] (B) -- (BB);
%\draw[dotted,->] (C) -- (CC);
%\draw[dotted,->] (DD) -- (BB);
%\draw[dotted,->] (E) -- (A);
%\draw[dotted,->] (D) -- (BB);

\draw[dotted,->] plot [smooth,tension=2] coordinates {(S6) (D)};
%\draw[dotted,->] plot [smooth,tension=1] coordinates {(S1) (.4,0) (B)};
%\draw[dotted,->] plot [smooth,tension=1] coordinates {(S3) (0,1) (CC)};
%\draw[dotted,->] plot [smooth,tension=1] coordinates {(S2) (-1,-1.4) (DD)};
%\draw[dotted,->] plot [smooth,tension=1] coordinates {(S5) (-2,-1) (A)};
%\draw[dotted,->] plot [smooth,tension=1] coordinates {(S7) (0,-1) (BB)};

\end{tikzpicture}\hspace{1cm}
\begin{tikzpicture}[scale=.8]
\node[nodevar] (A) at (0,0) {\var{1}};
\node[nodevar] (B) at (.5,-.5) {\var{8}};
\node[nodevar] (C) at (1,-1) {\var{7}};
\node[nodevar] (BB) at (-.5,-.5) {\var{2}};
\node[nodevar] (CC) at (-1,-1) {\var{3}};
\node[nodevar] (S1) at (0,.6) {\var{9}};
\node[nodevar] (S2) at (-1.6,-1) {\var{10}};
\node[nodevar] (S3) at (1,0) {\var{14}};

\draw[->] (B) -- (S3);
\draw[->] (A) -- (S1);
\draw[<-] (CC) -- (S2);
\draw[->] (A) -- (B);
\draw[->] (B) -- (C);
\draw[->] (A) -- (BB);
\draw[->] (BB) -- (CC);
\draw[color=red] (BB) circle (5pt);

\tikzset{every path/.style={color=black!30,shorten >=2pt}}
\node[nodevar] (E) at (0,-2) {\var{5}};
\node[nodevar] (DD) at (-.5,-1.5) {\var{4}};
\node[nodevar] (D) at (.5,-1.5) {\var{6}};
\node[nodevar] (S5) at (-1,-2) {\var{11}};
\node[nodevar] (S6) at (.5,-2.5) {\var{13}};
\node[nodevar] (S7) at (-.5,-2.5) {\var{12}};

\draw[->] (E) -- (D);
\draw[->] (D) -- (C);
\draw[->] (DD) -- (CC);
\draw[->] (E) -- (DD);

\draw[->] (E) -- (S6);
\draw[->] (S5) -- (DD);
\draw[->] (E) -- (S7);

\end{tikzpicture}
    \caption{The small, left, and top images of $q_{12}$. We do not draw the entire endomorphisms for readability.}
    \label{figure-qthree}
\end{figure}

\begin{example}\label{example-blowfish-counting}
The query $q_{12}$ can be enumerated with constant delay using its images depicted in \cref{figure-qthree}.
The idea is that we enumerate the solutions to the small image, and for every such assignment, we have a way to efficiently find all solutions that extend it. We first find all of its extensions to the top image and print the solutions implied by these. While we do this, we store the possible assignments to $x_7$. Then, we do the same with the left image and store the possible assignments to $x_{6}$. Next, We go over all pairs of value for $x_{6}$ and value for $x_7$ and check whether there is an edge between them. If there is an edge, we have found a non-trivial assignment to the cycle, and we can enumerate all possible spike assignments to get all solutions to the full query. The crux of this proof is that we can show that the number of pairs of values we need to check is bounded by the number of ``simple'' solutions we find using the two images. Thus, the ``simple'' solutions provide enough time to perform this computation while still achieving constant delay.
\end{example}

\begin{toappendix}
\vspace*{0.5\baselineskip}
{\bf Missing details for the query $q_{12}$ of \cref{example-blowfish-counting}}.
We first enumerate the solutions of the ``small'' image, depicted in \cref{figure-qthree}. For each such solution $(a,b,c,d,e)$, we do the following. 

We compute all possible assignments to $x_7$ while producing some solutions to $q_{12}$ using the top image as follows.
We refer to the subquery induced by $x_7,x_8,x_{14}$ as the top complement.
We enumerate the solutions to the top complement in which $x_8$ maps to $e$.
% to the query $p'$ depicted by \begin{tikzpicture}
% \node[fill=none] (A) at (0,0) {{$e$}};
% \node[nodevar] (B) at (.7,0) {\var{7}};
% \node[nodevar] (BB) at (.7,-.5) {\var{14}};
% \draw[->] (A) -- (B);
% \draw[->] (A) -- (BB);
% \end{tikzpicture} whose left value is $e$.
For each such solution $(j,e,k)$, we insert the value $j$ to the set $T_7(e)$. Notice that $(a,b,c,j,e,e,d,k)$ is a solution for the top image. We then output the solution to $q_{12}$ derived from this solution using the mapping of the top image. This is the solution $(a,b,c,b,a,e,j,e,e,d,a,b,e,k)$.

We continue similarly with the left image as follows.
We refer to the subquery induced by $x_4,x_5,x_6,x_{11},x_{12}$ as the left complement.
We enumerate the solutions to the left complement in which $x_4$ maps to $d$.
% of the query $p$ depicted by \begin{tikzpicture}
% \node[nodevar] (A) at (0,0) {{$d$}};
% \node[nodevar] (B) at (.7,0) {\var{5}};
% \node[nodevar] (C) at (1.2,0) {\var{6}};
% \node[nodevar] (BB) at (.7,-.3) {\var{11}};
% \node[nodevar] (CC) at (1.2,-.3) {\var{12}};
% \draw[->] (B) -- (A);
% \draw[->] (B) -- (C);
% \draw[->] (BB) -- (A);
% \draw[->] (B) -- (CC);
% \end{tikzpicture} whose left value is $d$.
For each such solution $(d,f,g,h,i)$, we insert the value $g$ to the set $T_6(d)$, and we insert the value $f$ to the set $T_5(d,g)$. Notice that $(a,b,c,d,f,g,e,d,h,i)$ is a solution for the left image. We then output the solution to $q_{12}$ derived from this solution using the mapping of the left image. This is the solution $(a,b,c,d,f,d,c,b,e,d,h,i,g,c)$.

Next, for each value $u$ in $T_6(d)$ and each value $v$ in $T_7(e)$ we check whether there is an edge $(u,v)$ in our database. If there is, for each value $w$ in $T_5(d,u)$, we have found a solution $(a,b,c,d,w,u,v,e)$ to the query induced by the cycle. We enumerate all possible spike assignments to get all solutions to the full query. More specifically, for each $s_1$ with an edge $(a,s_1)$, for each $s_2$ with an edge $(s_2,c)$, for each $s_3$ with an edge $(s_3,d)$, for each $s_4$ with an edge $(w,s_4)$, for each $s_5$ with an edge $(w,s_5)$, and for each $s_6$ with an edge $(e,s_6)$, we print the solution $(a,b,c,d,w,u,v,e,s_1,s_2,s_3,s_4,s_5,s_6)$ to $q_{12}$.

The process described in the last paragraph finds every solution to $q_{12}$ exactly once. The previous two paragraphs find some of the solutions to $q_{12}$, where the process described in each paragraph alone does not yield duplicate solutions. Overall, we find all solutions to $q_{12}$, and every solution is obtained at most $3$ times.

Let us now discuss the time complexity.
Each time we enumerate something in our algorithm, it is acyclic and can be done with linear preprocessing and constant delay.
The ``danger'' is that we need to go over all pairs of a value in $T_6(d)$ and a value in $T_7(e)$ and check whether they share an edge. In the worst case, if this check always fails, we can get a step of quadratic delay, which is a problem if we want linear preprocessing and constant delay.
However, we will show later that the number of such pairs is bounded by the number of solutions we print before these checks, using the top and left images. Overall, we get that, for all $n$, the time from the beginning of the execution until printing the $n$th solution is linear in the input size plus $n$. A general version of the Cheater's Lemma~\cite[Lemma 7]{UCQs} relies on this to deduce that our algorithm can be tweaked into working with linear preprocessing and constant delay with no duplicates, by withholding some of the solutions before printing them.

\newcommand\inset[1]{\text{In}({#1})}
\newcommand\outset[1]{\text{Out}({#1})}
\newcommand\solleft[1]{\text{SolLeft({#1})}}
\newcommand\soltop[1]{\text{SolTop({#1})}}
Denote by $\soltop{e}$ and $\solleft{d}$ the number of solutions printed in the second and third paragraphs respectively.
It is left to prove that $|T_6(d)|\cdot|T_7(e)|\le \soltop{e}+\solleft{d}$. 
Denote by $\inset{a}$ and $\outset{b}$ respectively the set of all values that go into $a$ and out of $b$ in the input database; that is $\inset{a}=\{v|(v,a)\in R\}$ and $\outset{b}=\{v|(b,v)\in R\}$. 
We have that $|T_7(e)|\le\outset{e}$ and $|T_6(d)|\le\sum_{f\in\inset{d}}{\outset{f}}$, and also $\soltop{e}=|\outset{e}|^2$ and $\solleft{d}=\sum_{f\in\inset{d}}{|\inset{d}||\outset{f}|^2}$.
Using the known inequality between the arithmetic and quadratic means, we get:
\begin{align*}
|T_6(d)|\cdot|T_7(e)|
\le |T_6(d)|^2 + |T_7(e)|^2
\le \left(\sum_{f\in\inset{d}}{\outset{f}}\right)^2 + \left(\outset{e}\right)^2\le |\inset{d}|\sum_{f\in\inset{d}}{\outset{f}}^2 + \outset{e}^2
=\soltop{e}+\solleft{d}
\end{align*}
This concludes the proof that $q_{12}$ can be enumerated with linear preprocessing and constant delay.

\end{toappendix}

The tractability that the spikes of $q_{12}$ introduce does not stem from the mere number of spikes. We next give evidence of the difficulty of enumerating the answers to the query $q_{13}$, which differs from $q_{12}$ only by the direction of one edge. 

\begin{example}\label{example-utd}
We can show that efficient enumeration for $q_{13}$ would imply a breakthrough in algorithms for triangle detection in unbalanced tripartite graphs, contradicting the following hypothesis.
The Vertex-Unbalanced Triangle Detection (\UTD) hypothesis\footnote{The \UTD hypothesis~\cite{UTDpaper} is less known than the previous ones used in this paper. However, it does represent a computational barrier, and relying on it gives evidence of the difficulty of the task at hand.} assumes that, for every constant $\alpha\in(0,1]$, it is not possible to test the existence of a triangle in a tripartite graph with vertex sets $|V|=n$ and $|U|=|W|=\Theta(n^\alpha)$ in time $O(n^{1+\alpha})$.
We encode triangle finding with tagging in the query like before, with two additional tricks.
To handle the two spikes leaving $x_5$, the value tagged by $x_5$ corresponds to an edge between $U$ and $V$. The same idea cannot be applied to $x_8$ as it would result in an image that contains vertices of the three parts that do not correspond to a triangle. The spike leaving $x_8$ makes one of the images correspond to two edges between $U$ and $W$. Here, we rely on the fact that our input graph is unbalanced to claim that there are at not too many solutions of this kind.
\end{example}
\begin{toappendix}

\vspace*{0.5\baselineskip}
{\bf Missing details for the query $q_{13}$ of \cref{example-utd}.}
Let $G$ be an unbalanced tripartite graph. We construct the following database $D$. For each node $u\in U$, we add to $D$ the facts $R(\pair{u,x_1},\pair{u,x_2})$,  $R(\pair{u,x_2},\pair{u,x_3})$, $R(\pair{u,x_4},\pair{u,x_3})$, $R(\pair{u,x_1},\pair{u,x_8})$, and $Red(\pair{u,x_2})$. For each edge $(u,v)\in U\times V$ of $G$, we add to $D$ the facts $R(\pair{(u,v),x_5},\pair{u,x_4})$ and $R(\pair{(u,v),x_5},\pair{v,x_6})$. For each edge $(v,w)\in V\times W$ of $G$, we add to $D$ the fact $R(\pair{(v,x_6},\pair{w,x_7})$. Finally for each edge $(w,u)\in W\times U$ of $G$, we add to $D$ the fact $R(\pair{u,x_8},\pair{w,x_7})$.
Consider now a solution in $q_{13}(D)$.
We perform a case analysis according to the tagging map, as explained in \cref{section-trick}, see \Cref{figure-blowfish-hard}.
The only images that contain $x_5$ or $x_6$ and also $x_7$ come from the endomorphisms which are the identity on the main loop, and so a solution with this tagging detects a triangle of $G$. Images that do not contain $x_5$, $x_6$, or $x_7$ result in solutions whose data part is a single node $u\in U$.
For images that contain $x_5$ and possibly also $x_6$ but not $x_7$, the data part is an edge $(u,v)$ between $U$ and $V$. For the image that contains $x_7$ and not $x_5$, the data part is a triple $(w,u,w')$ corresponding to two edges between $U$ and $W$. As $U$ and $W$ are small vertex sets, there cannot be too many solutions of the latter kind. 
Altogether the number of solutions that do not provide a triangle of $G$ is in $O(n^{1+\alpha})$. Hence, a constant delay algorithm after linear preprocessing time will identify a triangle in $G$ in time $O(n^{1+\alpha})$.

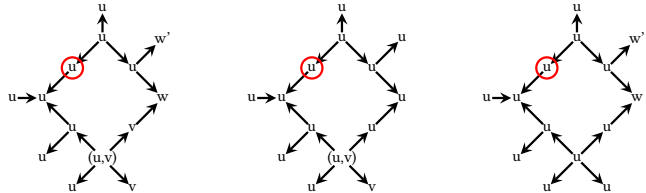
\begin{figure}[htb]
\centering
\begin{tikzpicture}[scale=.8]
\tikzset{mynodeAA/.style={color=white}}
\node[nodevar] (A) at (0,0) {\var{u}};
\node[nodevar] (B) at (.5,-.5) {\var{u}};
\node[nodevar] (C) at (1,-1) {\var{w}};
\node[nodevar] (D) at (.5,-1.5) {\var{v}};
\node[nodevar,rectangle] (E) at (0,-2) {\var{(u,v)}};
\node[nodevar] (BB) at (-.5,-.5) {\var{u}};
\node[nodevar] (CC) at (-1,-1) {\var{u}};
\node[nodevar] (DD) at (-.5,-1.5) {\var{u}};

\draw[->] (A) -- (B);
\draw[->] (B) -- (C);
\draw[->] (D) -- (C);
\draw[->] (E) -- (D);
\draw[->] (A) -- (BB);
\draw[->] (BB) -- (CC);
\draw[->] (DD) -- (CC);
\draw[->] (E) -- (DD);
\draw[color=red] (BB) circle (5pt);
\node[nodevar] (S1) at (0,.5) {\var{u}};
\node[nodevar] (S2) at (-1.5,-1) {\var{u}};
\node[nodevar] (S3) at (1,0) {\var{w'}};
\node[nodevar] (S5) at (-1,-2) {\var{u}};
\node[nodevar] (S6) at (.5,-2.5) {\var{v}};
\node[nodevar] (S7) at (-.5,-2.5) {\var{u}};

\draw[->] (A) -- (S1);
\draw[<-] (CC) -- (S2);
\draw[->] (B) -- (S3);
\draw[->] (DD) -- (S5);
\draw[->] (E) -- (S6);
\draw[->] (E) -- (S7);
\end{tikzpicture}\hspace{1cm}
\begin{tikzpicture}[scale=.8]
\tikzset{mynodeAA/.style={color=white}}
\node[nodevar] (A) at (0,0) {\var{u}};
\node[nodevar] (B) at (.5,-.5) {\var{u}};
\node[nodevar] (C) at (1,-1) {\var{u}};
\node[nodevar] (D) at (.5,-1.5) {\var{u}};
\node[nodevar,rectangle] (E) at (0,-2) {\var{(u,v)}};
\node[nodevar] (BB) at (-.5,-.5) {\var{u}};
\node[nodevar] (CC) at (-1,-1) {\var{u}};
\node[nodevar] (DD) at (-.5,-1.5) {\var{u}};

\draw[->] (A) -- (B);
\draw[->] (B) -- (C);
\draw[->] (D) -- (C);
\draw[->] (E) -- (D);
\draw[->] (A) -- (BB);
\draw[->] (BB) -- (CC);
\draw[->] (DD) -- (CC);
\draw[->] (E) -- (DD);
\draw[color=red] (BB) circle (5pt);
\node[nodevar] (S1) at (0,.5) {\var{u}};
\node[nodevar] (S2) at (-1.5,-1) {\var{u}};
\node[nodevar] (S3) at (1,0) {\var{u}};
\node[nodevar] (S5) at (-1,-2) {\var{u}};
\node[nodevar] (S6) at (.5,-2.5) {\var{v}};
\node[nodevar] (S7) at (-.5,-2.5) {\var{u}};

\draw[->] (A) -- (S1);
\draw[<-] (CC) -- (S2);
\draw[->] (B) -- (S3);
\draw[->] (DD) -- (S5);
\draw[->] (E) -- (S6);
\draw[->] (E) -- (S7);
\end{tikzpicture}\hspace{1cm}
\begin{tikzpicture}[scale=.8]
\tikzset{mynodeAA/.style={color=white}}
\node[nodevar] (A) at (0,0) {\var{u}};
\node[nodevar] (B) at (.5,-.5) {\var{u}};
\node[nodevar] (C) at (1,-1) {\var{w}};
\node[nodevar] (D) at (.5,-1.5) {\var{u}};
\node[nodevar,rectangle] (E) at (0,-2) {\var{u}};
\node[nodevar] (BB) at (-.5,-.5) {\var{u}};
\node[nodevar] (CC) at (-1,-1) {\var{u}};
\node[nodevar] (DD) at (-.5,-1.5) {\var{u}};

\draw[->] (A) -- (B);
\draw[->] (B) -- (C);
\draw[->] (D) -- (C);
\draw[->] (E) -- (D);
\draw[->] (A) -- (BB);
\draw[->] (BB) -- (CC);
\draw[->] (DD) -- (CC);
\draw[->] (E) -- (DD);
\draw[color=red] (BB) circle (5pt);
\node[nodevar] (S1) at (0,.5) {\var{u}};
\node[nodevar] (S2) at (-1.5,-1) {\var{u}};
\node[nodevar] (S3) at (1,0) {\var{w'}};
\node[nodevar] (S5) at (-1,-2) {\var{u}};
\node[nodevar] (S6) at (.5,-2.5) {\var{u}};
\node[nodevar] (S7) at (-.5,-2.5) {\var{u}};

\draw[->] (A) -- (S1);
\draw[<-] (CC) -- (S2);
\draw[->] (B) -- (S3);
\draw[->] (DD) -- (S5);
\draw[->] (E) -- (S6);
\draw[->] (E) -- (S7);
\end{tikzpicture}
\caption{Data parts of solutions in \Cref{example-utd} with a tagging map that: is the identity on the main loop (left), has $x_5$ in its image but not $x_7$ (center), and has $x_7$ in its image but not $x_5$ (right).}\label{figure-blowfish-hard} 
\end{figure}

\vspace*{0.5\baselineskip}
Notice that, for the construction of \Cref{example-utd}, it is important that the bottom-left spike points outwards. Otherwise, as in $q_{12}$, the left image may produce triplets of the form $(u,v_1,v_2)$ corresponding to two edges between $V$ and $U$, and there may be $n^{1+2\alpha}$ of them, which is too many for solving the \UTD problem in the desired time.

\end{toappendix}

\subsection{An open problem}\label{figure-open-cycle}

The techniques developed above for constant delay algorithms or hardness proofs fail on the full query $q_{14}$ depicted below, and its classification is left for future work. Note that $q_{14}$ is not a mirror, even though it consists of only one cycle and it has an acyclic image.
\begin{figure}[h]
\centering
\begin{tikzpicture}[baseline=0pt,rotate=90,scale=.8]
\node (A) at (0,0) {};
\node (B) at (0.5,0) {};
\node (C) at (1,0) {};
\node (D) at (1.2,-.5) {};
\node (E) at (1.4,-1) {};
\node (F) at (1,-1.2) {};
\node (G) at (1.4,-1.4) {};
\node (H) at (1.2,-1.9) {};
\node (I) at (1,-2.4) {};
\node (J) at (1.5,-2.6) {};
\node (K) at (1,-2.8) {};
\node (L) at (.5,-2.8) {};
\node (M) at (0,-2.8) {};
\node (N) at (-.5,-2.6) {};
\node (O) at (0,-2.4) {};
\node (P) at (-.2,-1.9) {};
\node (Q) at (-.4,-1.4) {};
\node (R) at (0,-1.2) {};
\node (S) at (-.4,-1) {};
\node (T) at (-.2,-.5) {};

\draw[->] (A) -- (B);
\draw[->] (B) -- (C);
\draw[<-] (C) -- (D);
\draw[<-] (D) -- (E);
\draw[->] (E) -- (F);
\draw[<-] (F) -- (G);
\draw[->] (G) -- (H);
\draw[->] (H) -- (I);
\draw[<-] (I) -- (J);
\draw[->] (J) -- (K);
\draw[<-] (K) -- (L);
\draw[<-] (L) -- (M);
\draw[->] (M) -- (N);
\draw[<-] (N) -- (O);
\draw[->] (O) -- (P);
\draw[->] (P) -- (Q);
\draw[<-] (Q) -- (R);
\draw[->] (R) -- (S);
\draw[<-] (S) -- (T);
\draw[<-] (T) -- (A);
\end{tikzpicture}
\end{figure}

\section{Conclusion}
We have initiated the study of the fine-grained complexity of conjunctive queries with self-joins.
We have seen that queries with low arity can be enumerated with constant delay iff their associated self-join-free query can. We do not know whether this property remains true with linear delay enumeration.
There are several ways in which the complexity analysis is more complicated for queries with self-joins. In particular, the complexity may change with the reordering of variables inside an atom, with the addition of unary atoms, or with the addition of binary atoms where only one of their variables appears in the rest of the query (called ``spikes'' above).
Despite showing several ways in which self-joins can be used in algorithms for queries that are otherwise hard, as well as showing ways to prove the hardness of queries with self-joins, determining the complexity of some queries (see \Cref{example-open} and \Cref{figure-open-cycle}) is left open.
Most of our results concern full conjunctive queries and going beyond this would require significant additional work.

\begin{acks}
We thank Louis Jachiet for his valuable input during our discussions.
This work was funded by the French government under management of Agence Nationale de la Recherche as part of the ``Investissements
d’avenir'' program, reference ANR-19-P3IA-0001 (PRAIRIE 3IA Institute), and by the
ANR project EQUUS ANR-19-CE48-0019.
\end{acks}

\bibliographystyle{ACM-Reference-Format}
\bibliography{main}

\end{document}